\documentclass[prb,twocolumn,nobibnotes]{revtex4-1}

\usepackage{amsmath}
\usepackage{mathtools}
\usepackage{amsfonts}
\usepackage{amssymb}
\usepackage{bm}
\usepackage{dsfont}

\usepackage{graphicx}
\usepackage{float}
\usepackage{microtype}

\usepackage{braket}
\usepackage{siunitx}
\usepackage{commath}

\usepackage[svgnames]{xcolor}
\usepackage[caption=false]{subfig}

\usepackage{hyperref}

\hypersetup{colorlinks,linkcolor={DarkBlue},citecolor={DarkBlue},urlcolor={DarkBlue}}
\usepackage[capitalize]{cleveref}

\newcommand{\R}[0]{\mathds{R}}
\newcommand{\wmax}[0]{\omega_{\text{max}}}

\newcommand{\dw}[0]{\dif \omega\;}
\newcommand{\e}[0]{\text{e}}
\newcommand{\ii}{\text{i}}
\renewcommand{\openone}{\mathds{1}}
\renewcommand{\epsilon}{\varepsilon}

\begin{document}

\title{Efficient simulation of open quantum systems coupled to a fermionic bath}

\author{Alexander N\"u{\ss}eler}
\affiliation{Institut f{\"u}r Theoretische Physik and Center for Integrated Quantum Science and Technology (IQST),\\
Albert-Einstein-Allee 11, Universit{\"a}t Ulm, 89069 Ulm, Germany}

\author{Ish Dhand}
\altaffiliation[Present address: ]{Xanadu, Toronto, Ontario, Canada M5G 2C8}
\affiliation{Institut f{\"u}r Theoretische Physik and Center for Integrated Quantum Science and Technology (IQST),\\
Albert-Einstein-Allee 11, Universit{\"a}t Ulm, 89069 Ulm, Germany}

\author{Susana F.~Huelga}
\affiliation{Institut f{\"u}r Theoretische Physik and Center for Integrated Quantum Science and Technology (IQST),\\
Albert-Einstein-Allee 11, Universit{\"a}t Ulm, 89069 Ulm, Germany}

\author{Martin B.~Plenio}
\affiliation{Institut f{\"u}r Theoretische Physik and Center for Integrated Quantum Science and Technology (IQST),\\
Albert-Einstein-Allee 11, Universit{\"a}t Ulm, 89069 Ulm, Germany}

\begin{abstract}
We present and analyze the fermionic time evolving density matrix using orthogonal polynomials algorithm (fTEDOPA), which enables the numerically exact simulation of open quantum systems coupled to a fermionic environment.
The method allows for simulating the time evolution of open quantum systems with arbitrary spectral densities at zero or finite temperatures with controllable and certified error.
We demonstrate the efficacy of the method towards the simulation of quintessential fermionic open quantum systems including the resonant level model and quantum dot coupled to an impurity and towards simulating hitherto intractable problems in quantum transport.
Furthermore, we demonstrate significant efficiency gains in the computational costs by performing simulations in the Heisenberg picture.
Finally, we compare different approaches for simulating finite-temperature situations and provide practical guidelines in order to identify the best strategy.  
\end{abstract}

\maketitle

\section{Introduction}

Ideal quantum systems may be considered closed, so that being isolated from their surrounding, they undergo textbook unitary evolution. 
In any real-world experimental setup, however, a given quantum system is open; that is, it interacts with an environment composed of those degrees of freedom that are not under the control of the experimenter~\cite{Breuer2002,Rivas2012,Alicki1987}. 
Hence, the numerical and analytical description of the dynamics of a quantum system in interaction with its environment is of fundamental importance in quantum physics. 

Some physically relevant situations, frequently encountered in quantum optics, involve systems that either couple weakly to an environment with vanishing memory time or are strictly confined to the high temperature limit.
Under these assumptions, it is possible to obtain an effective description of the system dynamics alone on the basics of additional assumptions that are physically motivated~\cite{Walls2008}. This microscopic approach typically leads to a reduced system dynamics that is expressed in terms of a Markovian master equation in Lindblad form~\cite{Lindblad1976, Gorini1976}.

In general, however, the interaction with structured or strongly coupled environments at finite temperatures can exhibit memory effects, so that the action of the environment on the system at a given time depends explicitly on the history of the system at all earlier times, giving rise to non-Markovian dynamics~\cite{Rivas2014,Breuer2016,Vega2017,Li2018}.
This will be the case for relevant problems encountered in physics, chemistry, and biology and for systems subject to either bosonic or fermionic reservoirs.
The bosonic case includes the spin-boson model important to solid state systems~\cite{Leggett1987} and biologically relevant molecular aggregates~\cite{Plenio2013}.
The interaction between a quantum system and its fermionic environment is central to the Kondo effect in quantum dots~\cite{Wiel2000,Sasaki2000}, molecular transistors~\citep{Etchegoin2002}, unconventional superconductors~\cite{Fritz2005}, and molecular magnets~\cite{Romeike2006}.

Lindblad master equations will generally fail to account for the system's evolution on those time scales where memory effects are significant and modeling the dynamics of these systems calls for more sophisticated methods.
In some circumstances, appropriate master equations may be derived provided the availability of an accurate, {\it ab-initio} Hamiltonian for the system-environment interaction~\cite{Hu1992,Shabani2005,Maniscalco2006}.
However, in many situations, and particularly when considering systems of increasing complexity, a microscopic derivation may just be unfeasible due to an insufficient knowledge of the system-environment interaction and/or the characteristics of the environmental spectral fluctuations.
In this case, modeling the dynamics of these systems calls for more sophisticated mathematical and numerical methods. 

Reaction coordinate mappings \cite{Garg1985,Garraway1997,Martinazzo2011,Woods2014} may provide a suitable strategy by incorporating the most relevant environmental modes into an extended system \cite{Nazir2018}.
However, when system extensions do not yield to a simple form for the residual spectral density of the environmental fluctuations, the accurate description of the system's dynamics may require the elimination of boundaries between system and reservoir and the deployment of numerical techniques aimed at the simulation of the global system.
Those "numerically exact" techniques include the use of hierarchical equations of motion~\cite{Tanimura1993}, iterative path integral methods~\cite{Makri1995a, Makri1995b, Segal2010}, numerical methods that use tensor network representations~\cite{Bulla2008, Schwarz2018} as well as combined techniques that utilize elements of both tensor networks coding and influence functional propagation~\cite{Strathearn2018}. 

Tensor network representations~\cite{Schollwoeck2011, Orus2014}, which underpin the availability of flexible and efficient numerical simulation methods with controllable and certified error, are central for the quantitative investigation of the dynamics of open quantum systems. 
In the case of quantum systems coupled to bosonic environments, one such method is provided by the time evolving density matrix using orthogonal polynomials algorithm (TEDOPA)~\cite{Chin2010,Prior2010}.
TEDOPA exploits tools from the theory of orthogonal polynomials and tensor networks for simulating the time-evolution of a quantum system coupled linearly to a bosonic or fermionic environment. 
Remarkably, TEDOPA allows for simulations of arbitrary spectral densities and coupling strengths with a controllable and certifiable error~\cite{Woods2015} and preserves full information about the state of the environment. 

TEDOPA has been employed to study dissipation in the spin-boson model~\cite{Chin2011} and time-frequency spectrum of the environmental excitations therein~\cite{Schroeder2016}; in the analysis of photonic crystals~\cite{Prior2013}; the study of polaron-polaritons in organic microcavities~\cite{DelPino2018}; the dynamics of populations and coherences in pigment-protein complexes~\cite{Chin2013}; and other biologically inspired quantum systems~\cite{Oviedo-Casado2016}.\\

The recently introduced method of thermalized TEDOPA~\cite{Tamascelli2019} reduces the computational cost of simulating finite-temperature situations by mapping the effect of finite-temperature thermal environments to zero temperature environments, thereby alleviating the costs associated with simulating mixed (thermal) states.
While TEDOPA is expected to simulate both bosonic and fermionic systems~\cite{Chin2010}, thus far the implementations and analysis have been restricted to the bosonic case.

The current work fills this gap by presenting fermionic TEDOPA (fTEDOPA) and analyzing its performance.
First, we analyze its application to the simulation of fermionic open quantum systems at zero temperature.
We underline the challenges in simulating fermionic environments, specifically those with finite chemical potential, which we overcome via fTEDOPA.
Furthermore, we demonstrate the potential of fTEDOPA simulations in the Heisenberg picture to increase the overall computational performance.
We also extend the results of finite-temperature simulations via thermalized TEDOPA~\cite{Tamascelli2019} from bosonic to fermionic environments and we highlight its strengths and limitations in comparison with competing approaches.
The structure of this paper is as follows.
\cref{Sec:Background} presents relevant background to place TEDOPA in the context of related methods.
In \cref{Sec:fermionic_tedopa}, we outline the steps involved in simulations via fTEDOPA and its thermalized version.
\cref{Sec:Numerics} presents numerical examples including the simulation of resonant level model and quantum dots.
We conclude with a discussion and open problems in \cref{Sec:Discussions}.
In order to keep this manuscript self-contained and as an invitation to newcomers in the field we have included detailed explanations in the main text.

\section{Background}
\label{Sec:Background}

In this section, we present the required background on bosonic TEDOPA~\cite{Chin2010,Prior2010}.
First, in \cref{Sec:setting}, we describe the systems that are in the purview of TEDOPA simulations, namely open quantum systems coupled linearly to a continuous environment of bosonic or fermionic nature.
TEDOPA uses ideas from the theory of orthogonal polynomials to map the continuous environment onto a semi-infinite chain of bosonic or fermionic modes with nearest-neighbor couplings.
We describe this so-called chain mapping in~\cref{Sec:ChainMapping}.
The chain thus obtained is simulated via methods from tensor networks, which we review in \cref{Sec:TNMethods}.

In \cref{Sec:TFTTBackground}, we review methods for simulating finite temperature environments via TEDOPA. 
We extend these methods to the fermionic regime in this work, as described in \cref{Sec:FermionicTFTT}.
We conclude the background section with a description of the numerical errors that arise from the assumptions made in TEDOPA in \cref{Sec:errors_tedopa}.

\subsection{The setting}
\label{Sec:setting}

We consider a system-environment Hamiltonian of the form
\begin{equation}
H = H_{\text{sys}} + H_{\text{env}} + H_{\text{int}},
\label{Eq:HFullOld}
\end{equation}
where the environment comprises bosonic or fermionic modes with frequencies $\omega$ and dispersion relation $g(\omega)$ described by the Hamiltonian
\begin{equation}
H_{\text{env}} := \int_{0}^{\omega_{\text{max}}} g(\omega)a_{\omega}^{\dagger}a_{\omega} \dif \omega.
\label{Eq:HEnv}
\end{equation}
Furthermore, the system-environment coupling is linear such that
\begin{equation}
H_{\text{int}} := \int_{0}^{\omega_{\text{max}}}h(\omega)(L^{\dagger} a_{\omega}+a^{\dagger}_{\omega} L) \dif \omega, \label{Eq:interaction_hamiltonian}
\end{equation}
for $L$ an arbitrary operator on the system and the real-valued coupling function $h(\omega)$ quantifying the strength of the coupling between each mode and the system.
In principle one could also think of a situation where the system couples to more than one bath.
However, for the sake of simplicity we will stick to the case of a single environment in this section. 

The environment is then described in terms of its spectral density 
\begin{equation}
J(\omega)=\pi h^{2}[g^{-1}(\omega)]\frac{\dif g^{-1}(\omega)}{\dif \omega},
\label{Eq:SpectralDensity}
\end{equation}
which captures the combined effect of the dispersion relation and the coupling function.
The factor ${\dif g^{-1}(\omega)}/{\dif \omega}$ describes the density of states of the environment in the frequency space and the factor $h^{2}[g^{-1}(\omega)]$ captures the strength of the interaction at a particular frequency.
Thus, the spectral density of \cref{Eq:SpectralDensity} describes the overall strength of interaction between the system and the environment modes as a function of the frequency.  
Moreover, given a spectral density $J(\omega)$, one can arbitrarily choose one of the two functions $g(\omega)$ and $h(\omega)$ if the other function is chosen such that \cref{Eq:SpectralDensity} holds.
One convenient choice for these functions for a given $J(\omega)$ is $g(\omega) = g\omega$, which gives 
\begin{equation}
h(\omega) = \sqrt{\frac{gJ(g\omega)}{\pi}},
\label{Eq:sd_coupling_linear}
\end{equation}
and we use this choice henceforth without loss of generality.

\subsection{Simulating system-environment time evolution via TEDOPA} \label{Sec:tedopa_discretization}

Here we provide relevant background about TEDOPA.
\cref{Sec:ChainMapping} describes the method of mapping the continuous star of \cref{Eq:interaction_hamiltonian} to a discrete chain, as performed by TEDOPA.
This chain is then simulated using tensor network methods that are detailed in \cref{Sec:TNMethods}.
For completeness, we provide a detailed exposition of these topics here and advise an expert reader to skip to \cref{Sec:fermionic_tedopa}.

\subsubsection{Mapping the continuous Hamiltonian to a discrete 1D Hamiltonian with nearest-neighbor couplings}
\label{Sec:ChainMapping}

\begin{figure}[htbp]
\centering
\includegraphics[width=\columnwidth]{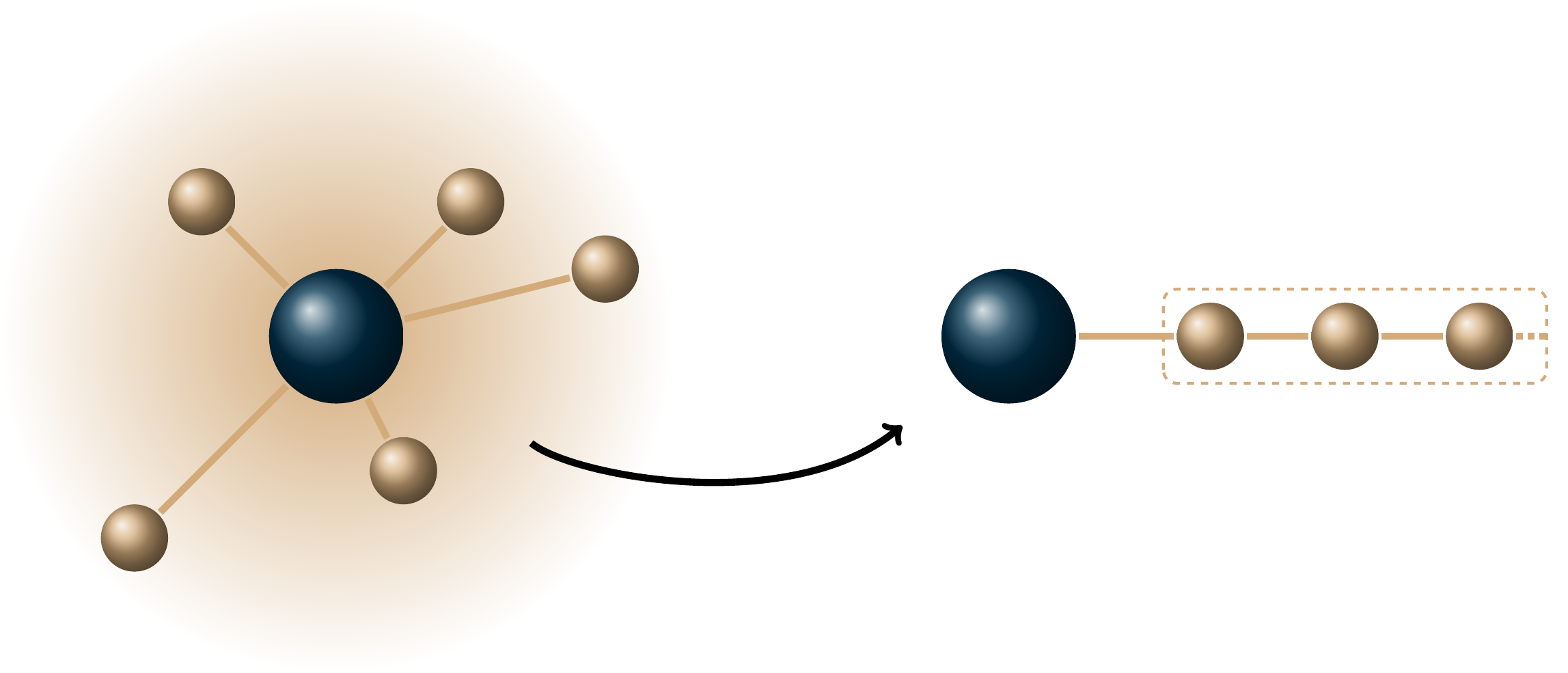}
\caption{\textbf{Chain mapping of a continuous bath.} 
In the left panel, the blue central dot represents the system and the shaded region represents the continuously coupled environment. 
The brown dots connected to the system represent the discretized environmental modes.
The right panel represents a system connected to a nearest-neighbor chain of virtual modes that are obtained directly from the continuous environment via the TEDOPA chain mapping step as detailed in the main text.
}
\label{Fig:ChainMapping}
\end{figure}

The full system-environment time evolution of \cref{Eq:HFullOld} is challenging to simulate in general as the environment comprises a continuous infinity of modes.
TEDOPA overcomes this challenge by mapping the continuous infinity to a discrete infinity of environmental modes.
Moreover, the initially free environment Hamiltonian $H_{\text{env}}$ of \cref{Eq:HEnv} is mapped to a Hamiltonian describing a semi-infinite chain of modes with nearest-neighbor couplings.
The system-environment interaction is mapped to a single coupling term between the system and the first mode of the chain.
This mapping is depicted in \cref{Fig:ChainMapping}.

In more detail, the first step of TEDOPA maps the bosonic or fermionic creation operators $a^{\dagger}_{\omega}$ to the new creation operators $b^{\dagger}_{n}$ as  
\begin{equation}
b_{n}^{\dagger} :=\int_{0}^{\wmax}\dif \omega\,U_{n}(\omega)a_{\omega}^{\dagger} \label{Eq:chain_mapping}
\end{equation}
where $n \in \mathbb{N}_0$ labels the discrete infinity of modes in the chain.
Choosing $U$ to be unitary, it can be shown that \cref{Eq:chain_mapping} preserves the canonical (anti-)commutation relations which means that the fermionic or bosonic nature of the environment is not altered under the mapping.
In particular, $U$ is constructed according to
\begin{equation}
U_n(\omega)=h(\omega)\frac{\pi_n(\omega)}{\|\pi_n\|}
\end{equation}
in terms of $\pi_{n}(\omega)$, the set of monic orthonormal polynomials with respect to the measure 
\begin{equation}
\dif \mu(\omega)=h^2(\omega)d\omega
\end{equation}
with support on $[0,\wmax]$.
Other choices of the measure have also been considered~\cite{Woods2015}.
Note that $h(x)$ is chosen such that the given spectral density is recovered by \cref{Eq:SpectralDensity} on setting $g(\omega) = g\omega$.  

This mapping clearly leaves the system Hamiltonian unchanged
\begin{equation}
H_{\text{sys}} \mapsto \tilde{H}_{\text{sys}} = H_{\text{sys}}.
\label{Eq:HSysNew}
\end{equation}
The interaction Hamiltonian transforms according to 
\begin{align}
H_{\text{int}} & \mapsto \tilde{H}_{\text{int}} \nonumber\\
& = \sum_n\frac{\|\pi_n\|^2\delta_{n0}}{\|\pi_n\|}(L^{\dagger} b_{n}+b_{n}^{\dagger} L)\nonumber\\
& = \|\pi_0\|  \;(L{^\dagger} b_{0}+ b_{0}^{\dagger} L)
\label{Eq:HIntNew}
\end{align}
and the environment Hamiltonian as
\begin{align}
&{H}_{\text{env}} \mapsto  \tilde{H}_{\text{env}}\nonumber\\
&= g\sum_{n}\left(\sqrt{\beta_{n+1}}b_{n}^{\dagger}b_{n+1}+\sqrt{\beta_{n+1}}b_{n+1}^{\dagger}b_{n}+\alpha_{n} b_{n}^{\dagger}b_{n}\right)
\label{Eq:HEnvNew}
\end{align}
in terms of $\alpha_{k}$ and $\beta_{k}$, which are the recursion coefficients 
\begin{equation}
\alpha_k=\frac{\langle x \pi_k,\pi_k\rangle}{\langle \pi_k,\pi_k\rangle}, \quad \beta_k=\frac{\langle \pi_k,\pi_k\rangle}{\langle \pi_{k-1},\pi_{k-1}\rangle}
\end{equation}
of the monic orthogonal polynomials~\cite{Chihara2011,Gautschi2004,Szego1967}.

The resulting Hamiltonians of \cref{Eq:HSysNew,Eq:HIntNew,Eq:HEnvNew} together provide the new Hamiltonian 
\begin{equation}
H_{\text{chain}} = {H}_{\text{sys}} + \tilde{H}_{\text{int}}  + \tilde{H}_{\text{env}}.
\label{Eq:Chain}
\end{equation}
Here, the Hamiltonian only comprises nearest-neighbor couplings among the environmental modes and the system couples only to the first of these modes.
Hence, we can interpret $H_{\text{chain}}$ as a semi-infinite chain Hamiltonian with the structure depicted in \cref{Fig:ChainMapping}.
This structure clearly resembles the output of the well-known chain-mapping introduced by Wilson~\cite{Bulla2008, Bulla2005, Wilson1975}.
However, in contrast to Wilson's originally proposed approach, the TEDOPA chain-mapping has the advantage of leading to unitarily equivalent representations and being numerically stable.
Moreover, it can be performed either analytically for some specific spectral densities~\cite{Chin2010} or numerically using stable algorithms~\cite{Gautschi1994}.
However, it differs conceptually in the sense that it aims to discretize the continuous bath with respect to the relevant features of its spectral density rather than with respect to the relevant energy scales.

\subsubsection{Tensor network methods}
\label{Sec:TNMethods}

The second step of TEDOPA concerns the simulation of the time evolution under the resulting nearest-neighbor Hamiltonian using tensor-network methods, such as matrix product states (MPSs) and operators (MPOs)~\cite{Schollwoeck2011}.
Here we recall relevant concepts of MPS and MPO methods. 

TEDOPA concerns the simulation of quantum states on $n$ subsystems, each of which is a $d$-level system.
These states are straightforwardly parameterized by coefficients $C_{i_{1}\dots i_{n}}$ according to 
\begin{equation}
\ket{\psi} = \sum_{i_{1}, \dots, i_{n} = 1}^{d} C_{i_{1}\dots i_{n}}\ket{i_{1}\dots i_{n}}
\label{Eq:Straightforward}
\end{equation}
for a suitable basis $\{\ket{i_{1}\dots i_{n}} \,|\, i_{1}, \dots, i_{n} = 1, \dots, d \}$.
This parameterization has $d^{n}$ elements, which makes computations infeasible for large $n$. 
MPSs replace \cref{Eq:Straightforward} with a product of matrices $\{A^{i_{1}}, A^{i_{2}}, \dots, A^{i_{n}} \}$ according to
\begin{equation}
\ket{\psi} = \sum_{\stackrel{i_{1}, \dots, i_{n}}{ = 1}}^{d}
\sum_{{j_{1}= 1}}^{D_{1}} 
\dots
\sum_{{j_{n-1}= 1}}^{D_{n-1}} 
A^{i_{1}}_{j_{1}}
A^{i_{2}}_{j_{1}j_{2}}
\dots
A^{i_{n}}_{j_{n-1}}
\ket{i_{1}\dots i_{n}}. \label{Eq:mps_representation}
\end{equation}
Here, $d$ is called the local dimension of the MPS.
The dimension $D_{k}$ of each of the matrices captures the amount of correlations across a bi-partition of system after the first k sites, i.e., comprising two halves: the first k subsystems $1, 2, \dots, k$ on the left and the remaining $k+1, \dots, n$ on the right. 
The maximum $D = \operatorname{max}_{k} D_{k}$ of these dimensions is referred to as the bond dimension of the MPS.
In combination with \cref{Eq:mps_representation} it becomes evident that the total number of parameters to be stored for an MPS is $\mathcal{O}(dD^2n)$.

MPSs provide an efficient parameterization of a set of relevant quantum states, including the ground states of local Hamiltonians, in which the entanglement and correlations are typically limited as they satisfy area laws~\cite{Eisert2010a}. 
These states are often considered as initial states for time evolution and can be well-approximated with MPSs whose bond dimension scales only polynomially with $n$ and can be stored and manipulated efficiently on a classical computer and their expectation values with respect to local measurements can be efficiently calculated.
Moreover, given a gapped local Hamiltonian in 1D, the MPS representation of its ground state can be found efficiently using the method of density matrix renormalization group (DMRG)~\cite{Schollwoeck2011,White1992a,Landau2015}.

In analogy to MPS representations of states, MPOs provide an efficient parameterization of a class of quantum operators.
This class includes density operators representing mixed states with low correlations and also unitary operators that capture time evolution under a local Hamiltonian for small times.
Consider an operator that has a canonical representation 
\begin{equation}
\hat{O} = 
\sum_{\stackrel{\stackrel{i_{1}, \dots, i_{n}}{i^{\prime}_{1}, \dots, i^{\prime}_{n}}}{ = 1}}^{d}
C_{i_{1}\dots i_{n}, i^{\prime}_{1}, \dots, i^{\prime}_{n}}\hat{\sigma}_{i_{1}i^{\prime}_{1}}\hat{\sigma}_{i_{2}i^{\prime}_{2}}\dots \hat{\sigma}_{i_{n}i^{\prime}_{n}}
\label{Eq:StraightforwardMPO}
\end{equation}
with an operator basis $\left\{\bigotimes_{k = 1}^{n}\hat{\sigma}_{i_k,i_k^{\prime}} \,|\, i_k,i_k^{\prime} = 1,\dots,d\right\}$.
The MPO representation of the operator is of the form 
\begin{align}
\begin{split}
\hat{O} =\,
\sum_{\stackrel{\stackrel{i_{1}, \dots, i_{n}}{i^{\prime}_{1}, \dots, i^{\prime}_{n}}}{ = 1}}^{d}
\sum_{{j_{1}= 1}}^{D_{1}} 
\dots
\sum_{{j_{n-1}= 1}}^{D_{n-1}}
& A^{i_{1},i^{\prime}_{1}}_{j_{1}}
\dots
A^{i_{n},i^{\prime}_{n}}_{j_{n-1}}
\hat{\sigma}_{i_{1}i^{\prime}_{1}}
\dots\hat{\sigma}_{i_{n}i^{\prime}_{n}}. \label{Eq:mpo_representation}
\end{split}
\end{align}
Here, each site $k$ has a basis labeled by two indices $i_{k}$ and $i^{\prime}_{k}$ indices so the effective local dimension is $d^{2}$, which leads to higher memory costs as compared to MPSs of comparable bond dimension.
In particular, the total number of parameters to be stored for an MPO is $\mathcal{O}(d^2D^2n)$.
Thus, MPO computation is more expensive and the cost is more sensitive to the local dimension of the system.

MPSs and MPOs enable the efficient simulation of $n$-partite quantum systems.
Simulating unitary evolution of pure states involves firstly storing the state as an MPS and then acting on this state with unitary operators returned by the Trotter-Suzuki decomposition~\cite{Trotter1959,Suzuki1976,Dhand2014a}.
The evolution of mixed states including finite-temperature thermal states is performed by storing the state density operator as an MPO and acting on this with unitary operators. 
Moreover, thermal states of given Hamiltonians can also be constructed using evolution in imaginary time starting from a maximally mixed state, which corresponds to cooling a state to the desired temperature.
A detailed exposition of MPS and MPO methods is presented in Ref.~\cite{Schollwoeck2011}.

\subsection{Thermalized and thermofield-based formulations}
\label{Sec:TFTTBackground}

Many physically relevant scenarios concern the dynamics of a system that is initially in a product state with its environment, which itself is considered to be in a thermal state at inverse temperature $\beta$.
Since thermal states are no longer pure states simulating such a configuration a priori requires the use of MPOs.
In general, this is expected to come along with increased computational costs which motivates the development of alternative simulation schemes that only rely on MPS methods.

A first thermofield-based alternative has been proposed by de Vega \textit{et al.}~\cite{Vega2015a}.
Consider again an open quantum system modeled by the Hamiltonian defined in \cref{Eq:HFullOld,Eq:HEnv,Eq:interaction_hamiltonian}.
For the ease of presentation we reparameterize the interaction Hamiltonian in terms of the spectral density $J(\omega)$ using \cref{Eq:SpectralDensity}.
The interaction Hamiltonian thus reads
\begin{align}
	H_{\text{int}} &= \frac{1}{\sqrt{\pi}} \int_{0}^{\wmax} \dw \sqrt{J(\omega)} \; (L^\dagger a_\omega + a_\omega^\dagger L).
	\label{Eq:interaction_hamiltonian_bosonic_thermofield}
\end{align}
The idea of the thermofield-based approach is to introduce a second ancillary bath and subsequently shift the temperature dependence of the initial state of the environment to the couplings between the system and the two baths.
This is done by performing a specifically chosen Bogoliubov transformation that maps the initial environment and interaction Hamiltonian to
\begin{align}
	\tilde{H}_{\text{env}} &= \int_{0}^{\wmax} \dw \omega \left(d_\omega^\dagger d_\omega - e_\omega^\dagger e_\omega \right)
\end{align}
and
\begin{align}
	\begin{split}
		\tilde{H}_{\text{int}}
		&= \frac{1}{\pi} \int_{0}^{\wmax} \dw \sqrt{J_1(\omega)} \left( L^\dagger d_\omega + d_\omega^\dagger L \right) \\
		&\hspace{15pt} + \frac{1}{\pi} \int_{0}^{\wmax} \dw \sqrt{J_2(\omega)} \left( L^\dagger e_\omega^\dagger + e_\omega L \right),
	\end{split}
\end{align}
respectively.
Here, $d_\omega$ and $e_\omega$ denote the transformed bosonic operators.
Moreover, the new spectral densities of the two baths are found to be~\cite{Vega2015a}
\begin{align}
	&J_1: [0, \wmax] \rightarrow \R, \; \omega \mapsto (1 + n_{\beta}(\omega)) J(\omega) \\
	&J_2: [0, \wmax] \rightarrow \R, \; \omega \mapsto n_{\beta}(\omega) J(\omega),
\end{align}
where $n_\beta$ denotes the Bose-Einstein distribution at inverse temperature $\beta$.
Interestingly, due to the thermal weighting of the initial spectral density, the thermal state of the initial Hamiltonian maps to the vacuum state of the transformed Hamiltonian.
Hence, we neither have to use an MPO to represent the initial state of the environment nor do we need to prepare it numerically.
A more detailed derivation of the thermofield-based approach for fermions is provided in Section~\ref{Sec:thermofield}, in which we also describe how to deal with finite chemical potential that is typical for fermionic environments.

Assuming a special form of the system-bath coupling Tamascelli \textit{et al.}~\cite{Tamascelli2019} have shown that two distinct baths are not necessarily needed to remove the temperature dependence from the initial state.
This second approach is referred to as thermalized TEDOPA and relies intimately on a factorized form of the system-bath interaction such that
\begin{align}
	H_{\text{int}} &:= L \; \frac{1}{\pi} \int_{0}^{\wmax} \dw \sqrt{J(\omega)} (a_\omega + a_\omega^\dagger).
\end{align}
Here, hermicity of the Hamiltonian is assured by assuming $L^\dagger = L$.
If system and environment additionally start off in a product state with the environment being in a Gaussian state, the dynamics of the system is solely determined by the first and second moments of the bath interaction operator at all times (see second paragraph of \cref{Sec:FTT_second_derivation} for more details).

This allows one to replace the spectral density $J$ as well as the initial state of the bath as long as the first and second moments are not altered.
In particular, the initial spectral density can be extended to the negative frequency domain to obtain the so-called thermalized spectral density $J_\beta: [-\wmax, \wmax] \rightarrow \R$ which is defined by
\begin{align}
	J_{\beta}(\omega) &:= \text{sign}(\omega) \frac{J(|\omega|)}{2} \left(1 + \coth\left(\frac{\beta\omega}{2}\right)\right).
\end{align}
Note that this extension to the negative frequency domain also changes the free environment Hamiltonian by introducing negative frequency modes very similar to the thermofield-based approach.
If one now chooses the initial state of the bath to be the vacuum state w.r.t. to this new Hamiltonian, one finds that the first and second moments of the bath interaction operators are equal for all times.
This means that the multi-time statistics of the two environments are the same and consequently the reduced dynamics of the system are identical.
Hence, thermalized TEDOPA, too, enables studying the system dynamics by only considering pure state evolution thus allowing the use of MPS instead of MPO evolution.
In Section~\ref{Sec:thermalized_tedopa} we show that the thermalized TEDOPA approach can also be generalized to the fermionic reservoirs.

\subsection{Errors in TEDOPA}
\label{Sec:errors_tedopa}

TEDOPA involves first mapping the continuous environment to a chain, and then using the MPS and MPO methods to perform time evolution on this chain.
Each of these steps introduces errors, which we describe systematically in this section.

In Section~\ref{Sec:ChainMapping}, we have seen that the TEDOPA chain mapping by itself is a unitary transformation and thus exact.
However, the one-dimensional chain of oscillators is not yet suited to perform numerical simulations using tensor network techniques.
This is due to the fact that firstly the set of discrete modes is still infinite (albeit countably infinite) and that the bosonic Fock space attached to each of the discrete modes is infinite dimensional.
We get rid of the first infinity by truncating the chain after a certain length $N$.
The value of $N$ sets the maximal simulation time you can achieve with good accuracy.
The second infinity is dealt with by truncating the local bosonic Hilbert space at a certain number of excitations, which is justified for many relevant applications, including those involving ground and low temperature thermal states.

The errors related to the TEDOPA chain mapping have been studied thoroughly in the past.
In particular, rigorous error bounds have been derived in the bosonic setting by Woods \textit{et al.}~\cite{Woods2015,Woods2016}.
Furthermore, the errors arising from finite $N$ have also been studied extensively numerically for paradigmatic scenarios in terms of the environment discretization by de Vega \textit{et al.}~\cite{Vega2015}.

A further source of error is due to the use of MPSs and MPOs.
For the sake of brevity, we will focus on MPS representations in the following but all arguments translate to MPOs as well.
Even though any state is exactly representable in MPS form, this requires exponentially large bond dimensions, which make simulations infeasible.
Hence, in all computations considered in this work, we follow usual MPS practice and allow for a certain approximation error such that 
\begin{align}
	\| \ket{\psi}_{\text{ex}} - \ket{\psi}_{\text{MPS}} \| < \epsilon.
\end{align}
In the MPS formalism this approximation error is controllable.
More precisely, given a state $\ket{\psi}_{\text{ex}}$ an approximation $\ket{\psi}_{\text{MPS}}$ is found by considering the Schmidt decomposition across all bipartitions of the chain.
Each Schmidt decomposition is then successively truncated either by discarding any Schmidt coefficient below a certain threshold or by only keeping a certain number of coefficients irrespective of their actual size.
While the number of retained coefficients determines the bond dimensions, the sum of all discarded coefficients yields the total approximation error $\epsilon$.
For a detailed treatment of this source of numerical errors, we refer to Section~4.5 of Schollw\"ock's review Ref.~\cite{Schollwoeck2011}.

Another essential part of TEDOPA is the simulation of the actual dynamics which suffers two sources of error that are common to MPS simulations~\cite{Schollwoeck2011}; firstly from imperfect preparation of initial states and secondly from imperfect evolution of real states in time.
For ground and thermal initial states, algorithms such as DMRG or imaginary time evolution are used to construct the MPS representation of these states.
Naturally, both DMRG and imaginary time evolution suffer errors related to the MPS approximation of the state.
Additionally, DMRG incurs errors from the local optimization as well as the total number of optimization steps.
On the other hand, imaginary time evolution is subject to errors stemming from the propagator in time, i.e., errors that depend on the step-width $\delta t$ and the Suzuki-Trotter expansion itself.
Once the initial state is prepared, real time evolution is performed on the system which suffers from the same error sources as imaginary time evolution.

\section{Fermionic TEDOPA}
\label{Sec:fermionic_tedopa}

In this section, we present the method of fTEDOPA.
We focus first on the two key challenges in formulating fTEDOPA that are different from bosonic TEDOPA.
Secondly, we provide proofs for the generalization of thermalized TEDOPA~\cite{Tamascelli2019} to the fermionic setting.
The first step of TEDOPA, namely the chain-mapping step, is formally identical to the bosonic case as already described in literature~\cite{Chin2010}.
The second step, that of simulating the system-chain interaction for a fermionic chain diverges from the bosonic situation in two important respects.
The first issue is that of accounting for the anti-symmetry properties of the wave function of the obtained chain modes and also of the system modes if the system comprises fermions of the same species as the chain.
The second issue concerns the presence of negative frequency normal modes of the fermionic chain which corresponds to a positive chemical potential and is unique to the fermionic case~\cite{Rau2017}.
We address these two issues in the remainder of this section.

\subsection{Jordan-Wigner transformation}
\label{Sec:jordan_wigner_transformation}

Firstly, we consider the issue of the anti-symmetric wave function of a fermionic system.
Suppose we denote the set of creation operators by $\lbrace f_k^\dagger \rbrace_{k=0}^N$.
The anti-symmetry property then corresponds to the fermionic creation and annihilation operators obeying the canonical anti-commutation relations (CAR), i.e.,
\begin{align}
	\{f_k, f_\ell\} = \{f_k^\dagger,f_\ell^\dagger\} = 0 \quad \{f_k, f_\ell^\dagger \} = \delta_{k,\ell}. 	\label{Eq:fermionic_car}
\end{align}
While for bosons there is a straightforward matrix representation of creation and annihilation operators in the Fock basis, obtaining such matrix representations for fermions is more involved.
To solve this problem we use the Jordan-Wigner transformation which maps the fermionic chain onto a chain of spins.
We start by fixing the order of the fermionic operators to uniquely define the Fock space.
Here, we follow the convention
\begin{align}
    |n_1, n_2,\dots \rangle &:= (f_1^\dagger)^{n_1}(f_2^\dagger)^{n_2}\dots(f_N^\dagger)^{n_N} |\Omega\rangle \label{Eq:fock_space_order}
\end{align}
where $|\Omega\rangle$ is the vacuum state with respect to the fermionic operators $\left\{ f_k : k = 1,\dots,N \right \}$, i.e., $f_k |\Omega\rangle = 0$ for any $k \in \{1,\dots,N\}$. 
Having fixed this specific order of operators the Jordan-Wigner transformation maps fermionic operators to spin operators according to~\cite{Nielsen2005}
\begin{align}
    f_k &\mapsto \left( \bigotimes_{\ell = 1}^{k-1} \sigma^z_\ell \right) \otimes \sigma^-_k, & f_k^\dagger &\mapsto \left( \bigotimes_{\ell = 1}^{k-1} \sigma^z_\ell \right) \otimes \sigma^+_k, \label{Eq:jordan_wigner_transformation}
\end{align}
where
\begin{align}
    \sigma_k^+ &:= 
    \begin{pmatrix}
        0 & 1 \\ 0 & 0
    \end{pmatrix} &
    \sigma_k^- &:= 
    \begin{pmatrix}
        0 & 0 \\ 1 & 0
    \end{pmatrix}& 
    \sigma_k^z &:= 
    \begin{pmatrix}
        1 & 0 \\ 0 & -1
    \end{pmatrix} \label{Eq:spin_matrices}
\end{align}
are the usual spin operators.
The form of the transformation in \cref{Eq:jordan_wigner_transformation} together with the intrinsic algebra of the $2\times2$ spin matrices in \cref{Eq:spin_matrices} ensure that the fermionic CAR in \cref{Eq:fermionic_car} are satisfied.

We apply the Jordan-Wigner transformation to Hamiltonians of the form of \cref{Eq:HEnvNew}.
These Hamiltonians comprise terms that are quadratic in fermionic operators and these terms can be mapped to spin operators as follows. 
First, we consider number operators, which have the form $n_k = f_k^\dagger f_k$.
Applying \cref{Eq:jordan_wigner_transformation} and exploiting $(\sigma_\ell^z)^2 = \openone$, the spin representation of the number operator is
\begin{align}
    n_k = f_k^\dagger f_k &= \left( \bigotimes_{\ell = 1}^{k-1} (\sigma_\ell^z)^2 \right) \otimes \sigma_k^+ \sigma_k^- = \sigma_k^+ \sigma_k^-. \label{Eq:numop_jordan_wigner}
\end{align}
Next we consider a general jump operator $f_n^\dagger f_m$ with $n < m$, which transforms according to
\begin{align}
    f_n^\dagger f_m &= \left(\bigotimes_{\ell = 1}^{n-1} (\sigma_\ell^z)^2 \right) \otimes \sigma_n^+ \sigma_n^z \otimes \left(\bigotimes_{\ell = n+1}^{m-1} \sigma_\ell^z \right) \otimes \sigma_m^-  \\
    &= - \sigma_n^+ \otimes \left(\bigotimes_{\ell = n+1}^{m-1} \sigma_\ell^z \right) \otimes \sigma_m^-, \label{Eq:non_local_jump_operator}
\end{align}
where going from the first to the second line we used $\sigma_k^+ \sigma_k^z = -\sigma_k^+$.

While \cref{Eq:non_local_jump_operator} reveals that a general non-local jump operator, in the sense of the order imposed by \cref{Eq:fock_space_order}, is mapped to a non-local operator on the spin-chain, for  nearest-neighbor interactions $f_n^\dagger f_{n+1}$ the expression simplifies to
\begin{align}
    f_n^\dagger f_{n+1} &= - \sigma_n^+ \otimes \sigma_{n+1}^-. \label{Eq:hopping_jordan_wigner1}
\end{align}

In fTEDOPA we consider two situations: either a system that includes fermionic modes of the same species as the chain or not.
In the former case the system is subject to the anti-symmetry constraint implemented by the CAR and thus has to be transformed via the Jordan-Wigner transformation.
Let us point out two examples we will pick up again in Section~\ref{Sec:Numerics}.

The first example is a single fermionic mode $f_0$ that is linearly coupled to a nearest-neighbor chain of other fermionic modes $\lbrace f_k \rbrace_{k = 1}^\infty$ of the same species.
Specifically, we consider a Hamiltonian 
\begin{align}
    \begin{split}
        H &= \omega_0 f_0^\dagger f_0 + t_1 (f_0^\dagger f_1 + f_1^\dagger f_0)\\
        &\hspace{15pt} + \sum_{k=1}^\infty \omega_k f_k^\dagger f_k + \sum_{k=2}^\infty t_k \left(f_\text{k}^\dagger f_{k+1} + f_{k+1}^\dagger f_k\right)
    \end{split}
\end{align}
where $\{\omega_k\}_{k=0}^\infty$ determines the energies and $\{t_k\}_{k=1}^\infty$ the couplings of the modes.
Using \cref{Eq:numop_jordan_wigner} and \eqref{Eq:hopping_jordan_wigner1}, the fermionic Hamiltonian maps to the spin Hamiltonian
\begin{align}
    \begin{split}
        \tilde{H} &= \omega_0 \sigma_0^+ \sigma_0^- + \sum_{k=1}^\infty \omega_k \sigma_k^+ \sigma_k^- \\
        &\hspace{15pt} - t_1 (\sigma_0^+ \otimes \sigma_1^- + \sigma_0^- \otimes \sigma_1^+) \\
        &\hspace{15pt} - \sum_{k = 2}^\infty t_k (\sigma_{k}^+ \otimes \sigma_{k+1}^- + \sigma_{k}^- \otimes \sigma_{k+1}^+).
    \end{split}
\end{align}

An example of the distinguishable setting is a popular model of quantum dots which extends the previous example by coupling another two-level system to the single fermionic mode $f_0$.
This leads to the Hamiltonian
\begin{align}
	\begin{split}
        H &= \omega_{\text{tls}} \sigma_z + A_{\text{tls}} \otimes n_0 + \omega_0 f_0^\dagger f_0  + t_1 (f_0^\dagger f_1 + f_1^\dagger f_0)\\
        &\hspace{15pt} + \sum_{k=1}^\infty \omega_k f_k^\dagger f_k + \sum_{k=2}^\infty t_k \left(f_k^\dagger f_{k+1} + f_{k+1}^\dagger f_k\right).
	\end{split}
\end{align}
Here, $\omega_{\text{tls}}$ and $A_{\text{tls}}$ denote the energy separation and the coupling operator of the two-level system, respectively.
Applying the Jordan-Wigner transformation finally yields
\begin{align}
	\begin{split}
		\tilde{H} &= \omega_{\text{tls}} \sigma_z + A_{\text{tls}} \otimes \sigma_0^+ \sigma_0^- \\
        &\hspace{10pt} + \omega_0 \sigma_0^+ \sigma_0^- - t_1 (\sigma_0^+ \otimes \sigma_1^- + \sigma_0^- \otimes \sigma_1^+) \\
        &\hspace{10pt} + \sum_{k=1}^\infty \omega_k \sigma_k^+ \sigma_k^- - \sum_{k = 2}^\infty t_k (\sigma_{k}^+ \otimes \sigma_{k+1}^- + \sigma_{k}^- \otimes \sigma_{k+1}^+). 
	\end{split}
\end{align}

In summary, by applying the Jordan-Wigner transformation in \cref{Eq:jordan_wigner_transformation} to the fermionic Hamiltonians we obtain a corresponding spin Hamiltonian that ensures the fermionic CAR in \cref{Eq:fermionic_car} and is well suited for the use of tensor network techniques.
In particular, the nearest-neighbor chain Hamiltonians conveniently transform to nearest-neighbor spin Hamiltonians which allow for the use of specialized time-evolution algorithms such as time evolving block decimation (TEBD)~\cite{Vidal2003, Vidal2004}.

\subsection{State preparation}
\label{Sec:state_preparation}

Another important factor to account for in simulations is the preparation of the MPS representation of the initial state.
Usual applications involve situations where the initial state is a thermal or ground state of the environment Hamiltonian $H_{\text{env}}$ or the full Hamiltonian $H_{\text{chain}}$ which is then evolved under a quenched Hamiltonian.
Constructing thermal states of two Hamiltonians $H_{\text{env}}$ and $H_{\text{chain}}$ correspond to two different situations---the system being either coupled to or decoupled from the environment before the start of the time evolution.
Below we elaborate on the special features of fermionic state preparation in both these situations, highlighting their difference to initialization with bosonic modes.

The simplest state preparation scenario is in the bosonic case when the initial state is a factorized state between system and environment at zero temperature.
This case involves a ground state of the star Hamiltonian \eqref{Eq:HEnv}.
This Hamiltonian is already decomposed in its eigenbasis, formed by the so-called normal modes. 
In the bosonic setting the chemical potential is either zero or negative~\cite{[See Sec. 6.1 of ]Rau2017}.
Thus, the ground state is given by the vacuum state, i.e., $a_\omega |\Omega\rangle = 0$ for all $\omega \in [0, \omega_{\max}]$.
TEDOPA requires transforming this state into the chain picture. 
Fortunately, applying the chain transformation defined in \cref{Eq:chain_mapping} simply maps the vacuum state w.r.t. the operators $a_\omega$ to the vacuum state w.r.t. the operators $b_\omega$.
This vacuum state is convenient for numerics since it is an uncorrelated product state and can thus be trivially constructed as an MPS with all bond dimensions equal to one.

In contrast to bosons, fermions allow for a positive chemical potential which means that the ground state is no longer the vacuum state.
It is rather the state where all normal modes with energy below the chemical potential are fully occupied and all modes with energy above the chemical potential are unoccupied.
While this is still a uncorrelated product state in the star geometry, applying the chain mapping and the Jordan-Wigner transformation could potentially yield a highly correlated state since all the normal modes are mixed.
This fermionic peculiarity has two implications. 
Firstly, the possibly increased amount of correlations in the state leads to higher bond dimensions in the MPS representation.
Secondly, we are not able to write down the MPS representation straightforwardly in general.
Hence, we have to resort to tensor network algorithms such as imaginary time evolution or variational DMRG to find the ground state~\cite{Schollwoeck2011}.
This aspect of the fermionic state preparation can severely increase the overall computational effort.

The increased amount of correlations in the fermionic ground state also carries over to low temperature thermal states.
A general thermal state at inverse temperature $\beta$ is prepared by evolving the totally mixed state $\rho = \frac{1}{\mathcal{Z}} \openone$ in imaginary time which effectively cools the system down to the desired $\beta$.
As in the case of ground states discussed before, a positive chemical potential induces more excitations a priori and thus in general increases the overall amount of correlations in the thermal states of the chain.

For high temperature, i.e., low $\beta$, the situation changes.
While a single boson has an infinite-dimensional Fock space, the dimension of the Fock space of a single fermion is equal to two.
Assuming for the moment we were able to work with infinite-dimensional Hilbert spaces, both, fermionic and bosonic, thermal states would approach the totally mixed state as $\beta$ goes to zero. 
However, for bosons the infinite-dimensional Fock space has to be truncated at a certain maximal occupation number in order to apply tensor network techniques.
The choice of this truncation threshold strongly depends on the value of $\beta$.
This is because Bose-Einstein statistics tells us that with increasing temperature the occupation of the normal modes also increases.
Hence, keeping the approximation error constant requires truncating the Fock space at higher occupation numbers.
Consequently, the memory complexity of a bosonic thermal state has to diverge for $\beta$ going to zero.
For fermions, however, there is no truncation of the Fock space involved which means that representing a fermionic thermal state at high temperature becomes very efficient.

\subsection{Errors in fTEDOPA}

As we have seen in Section~\ref{Sec:errors_tedopa}, there are several sources of errors involved in TEDOPA.
These errors are influenced by the fermionic nature of the open quantum system in both ways, positive and negative.
On the one hand, fTEDOPA does not suffer from errors stemming from the truncation of the local Hilbert space.
This is because the fermionic Fock space is finite dimensional.
On the other hand, as we argued in the previous section, a large set of physically relevant, fermionic states suffers an increased amount of correlations compared to bosons.
This reflects especially in the preparation of initial states but also in subsequent time evolution.
In particular, being limited by the available memory resources, the truncation of bond dimensions has to be more crude, i.e., can involve a larger truncation error.
These larger errors in the initial state can also lead to larger errors in the subsequent evolution.

\section{Finite-temperature fermionic TEDOPA}
\label{Sec:FermionicTFTT}

Having emphasized in Section~\ref{Sec:state_preparation} that thermal states can lead to substantially harder simulation problems, partially due to the necessity of MPO representations, this section presents alternatives that only require the use of MPSs.
In particular, we firstly revise the thermofield-based approach~\cite{Vega2017}.
Subsequently, we outline two approaches to extend thermalized TEDOPA~\cite{Tamascelli2019} to the fermionic setting.
While the first approach is based on the thermofield approach, the second relies on directly replacing the spectral density of the environment, following closely the derivation presented in Ref.~\cite{Tamascelli2019}.
Nonetheless the two approaches provide the same discretized chain in the end.

\subsection{Thermofield approach}
\label{Sec:thermofield}

Consider again a global Hamiltonian of the form
\begin{align}
	\begin{split}
		H :=\,& H_{\text{sys}} + H_{\text{env}} + H_{\text{int}}, \\
		H_{\text{env}} :=\,& \int_{0}^{\wmax} \dw \omega f_\omega^\dagger f_\omega, \\
		H_{\text{int}} :=\,& \int_{0}^{\wmax} \dw h(\omega) \; (L^\dagger f_\omega + f_\omega^\dagger L),
	\end{split}
\end{align}
where $H_{\text{sys}}$ denotes the free Hamiltonian of the system and $L$ its coupling operator.

Firstly, let us introduce a second, non-interacting ancillary bath with fermionic operators $g_\omega$ of the same species,
\begin{align*}
H_{\text{anc}} &:= \int_{0}^{\wmax} \dw \omega g_\omega^\dagger g_\omega,
\end{align*}
such that
\begin{align}
	\begin{split}
		\tilde{H} :=\,& H_{\text{sys}} + \tilde{H}_{\text{env}} + H_{\text{int}} \\
		\tilde{H}_{\text{env}} :=\,& H_{\text{env}} - H_{\text{anc}}.
	\end{split}
\end{align}
This new bath is not coupled to the system via $H_{\text{int}}$ and thus the dynamics remain unchanged.
Secondly, we apply a fermionic Bogoliubov transformation on these two sets of operators, i.e., we will define new fermionic operators $d_\omega$ and $e_\omega$ which are related to the former ones by
\begin{align}
	\begin{split}
		f_\omega &= u(\omega) d_\omega + v(\omega) e_\omega^\dagger, \quad f_\omega^\dagger = u(\omega) d_\omega^\dagger + v(\omega) e_\omega, \\
		g_\omega &= u(\omega) e_\omega - v(\omega) d_\omega^\dagger, \quad g_\omega^\dagger = u(\omega) e_\omega^\dagger - v(\omega) d_\omega.
	\end{split}\label{Eq:bogtransform}
\end{align}
Here $u, v: [0, \wmax] \rightarrow \R$ and we require
\begin{equation}
u(\omega)^2 + v(\omega)^2 = 1 \qquad \forall \omega \in [0, \wmax]
\label{Eq:cossinbogo}
\end{equation}
in order to preserve the fermionic nature of the operators, i.e., the fermionic CAR.

Given this transformation we can compute the number operators in terms of the new basis, which leads to 
\begin{align}
		f_\omega^\dagger f_\omega &= (u(\omega) d_\omega^\dagger + v(\omega) e_\omega) (u(\omega) d_\omega + v(\omega) e_\omega^\dagger) \nonumber\\
	\begin{split}
		&= u(\omega)^2 d_\omega^\dagger d_\omega + u(\omega) v(\omega) d_\omega^\dagger  e_\omega^\dagger \\
		&\hspace{15pt} + u(\omega) v(\omega) e_\omega d_\omega +  v(\omega)^2 e_\omega e_\omega^\dagger
	\end{split}\nonumber\\
	\begin{split}
		&=  u(\omega)^2 d_\omega^\dagger d_\omega - v(\omega)^2 e_\omega^\dagger e_\omega + u(\omega) v(\omega) d_\omega^\dagger e_\omega^\dagger\\
		&\hspace{15pt}+ v(\omega) u(\omega) e_\omega d_\omega + v(\omega)^2,
	\end{split}\label{Eq:fkdagfk}
\end{align}
where we have used fermionic CAR to bring the expression into normal ordering in the last line. 
Likewise, the number operator of the second bath is expressed in terms of the new operators according to 
\begin{align}
		g_\omega^\dagger g_\omega &= (u(\omega) e_\omega^\dagger - v(\omega) d_\omega)(u(\omega) e_\omega - v(\omega) d_\omega^\dagger)\nonumber\\
	\begin{split}
		&= u(\omega)^2 e_\omega^\dagger e_\omega - v(\omega)^2 d_\omega^\dagger d_\omega \\
		&\hspace{15pt} - u(\omega)v(\omega) e_\omega^\dagger d_\omega^\dagger - v(\omega)u(\omega) d_\omega e_\omega + v(\omega)^2 
	\end{split}\nonumber\\
	\begin{split}
		&= u(\omega)^2 e_\omega^\dagger e_\omega - v(\omega)^2 d_\omega^\dagger d_\omega + u(\omega)v(\omega) d_\omega^\dagger e_\omega^\dagger \\
		&\hspace{15pt} + v(\omega)u(\omega) e_\omega d_\omega + v(\omega)^2.
	\end{split} \label{Eq:gkdaggk}
\end{align}
Again the CAR are used to bring the expression into the same ordering as \cref{Eq:fkdagfk}.

Hence, the free Hamiltonian of the two baths maps to
\begin{align}
	\tilde{H}_{\text{env}} &\mapsto \tilde{H}^\prime_{\text{env}} = \int_{0}^{\wmax} \dw \omega (f_\omega^\dagger f_\omega - g_\omega^\dagger g_\omega) \nonumber\\
	\begin{split}
		&= \int_{0}^{\wmax} \dw \omega (u(\omega)^2 + v(\omega)^2) d_\omega^\dagger d_\omega \\
		&\hspace{10pt}- \int_{0}^{\wmax} \dw \omega (u(\omega)^2 + v(\omega)^2) e_\omega^\dagger e_\omega.
	\end{split}
\end{align}
Exploiting \cref{Eq:cossinbogo} this simplifies to
\begin{align}
	\tilde{H}^\prime_{\text{env}} &= \int_{0}^{\wmax} \dw \omega \left(d_\omega^\dagger d_\omega - e_\omega^\dagger e_\omega \right). \label{eq:btraf_free_hamiltonian_discrete1}
\end{align}
For the interaction part, we obtain
\begin{align}
	\begin{split}
		H_{\text{int}} &\mapsto H^\prime_{\text{int}} = \int_{0}^{\wmax} \dw  h(\omega) L^\dagger (u(\omega) d_\omega + v(\omega) e_\omega^\dagger) \\
		&\hspace{15pt} + \int_{0}^{\wmax} \dw  h(\omega) (u(\omega) d_\omega^\dagger + v(\omega) e_\omega) L
	\end{split}\nonumber\\
	\begin{split}
		&= \int_{0}^{\wmax} \dw h(\omega) u(\omega) \left( L^\dagger d_\omega + d_\omega^\dagger L \right) \\
		&\hspace{15pt} + \int_{0}^{\wmax} \dw h(\omega) v(\omega) \left( L^\dagger e_\omega^\dagger + e_\omega L \right).
	\end{split}
 \label{eq:btraf_interaction_hamiltonian}
\end{align}

The reason to change the bases according to \cref{Eq:fkdagfk} is as follows.
In order to see the fundamental difference between the initial and the transformed Hamiltonian, consider a thermal state of the free environment in the initial operator basis at inverse temperature $\beta$ and chemical potential $\mu$,
\begin{align}
	\rho_{\beta,\mu}^{\text{env}} := \frac{\e^{-\beta (H_{\text{env}} - \mu N)}}{\mathcal{Z}}. \label{Eq:thermal_state_chemical_potential}
\end{align}
Here, $\mathcal{Z}$ denotes the partition function and
\begin{equation}
N := \int_0^{\wmax}\dw f_\omega^\dagger f_\omega,
\end{equation}
which is the global number operator.
Furthermore, we assume the ancillary bath to be in a thermal state $\rho_{\beta, \mu}^{\text{anc}}$ with the same inverse temperature $\beta$ and the same chemical potential $\mu$.
For a non-interacting system of fermionic modes the expectation values of the number operators in the thermal state follow the Fermi-Dirac distribution, i.e., 
\begin{align}
	\hspace{-5pt}\langle f_\omega^\dagger f_\omega \rangle_{\rho_{\beta,\mu}^{\text{env}}}  &= \langle g_\omega^\dagger g_\omega \rangle_{\rho_{\beta,\mu}^{\text{anc}}} =: \bar{n}_{\beta, \mu}(\omega) = \frac{1}{\e^{\beta (\omega - \mu)} + 1}.
\end{align}
In contrast, taking the expectation value in the vacuum state of the environment w.r.t. the transformed fermionic modes, $\rho_\Omega := |\Omega\rangle \langle\Omega|$, we obtain via \cref{Eq:fkdagfk} and \eqref{Eq:gkdaggk}
\begin{align}
	\langle f_\omega^\dagger f_\omega \rangle_{\rho_\Omega} &= \langle g_\omega^\dagger g_\omega \rangle_{\rho_\Omega} = v(\omega)^2.
\end{align}
Hence, by choosing
\begin{align}
	v: [0, \wmax] \rightarrow [0,1], \; \omega \mapsto \sqrt{\bar{n}_{\beta,\mu}(\omega)}, \label{Eq:thermofield_v_function}
\end{align}
we fix the transformation in \cref{Eq:bogtransform} such that the thermal state of the environment w.r.t. the initial modes $\rho_\beta$ is mapped to the vacuum state w.r.t. the transformed modes $\rho_\Omega$.
Note that \cref{Eq:thermofield_v_function} implicitly defines
\begin{align}
	u: [0, \wmax] \rightarrow [0,1], \; \omega \mapsto \sqrt{1 - \bar{n}_{\beta,\mu}(\omega)}
\end{align}
due to the constraint in \cref{Eq:cossinbogo}.
Since the vacuum state $|\Omega\rangle$ is a pure state, the potential advantage of the transformed Hamiltonian becomes more evident.
While in the initial setting we have to use an MPO to represent the thermal state of the environment, now an MPS is sufficient.
However, even though the usage of MPSs seems to be favorable in terms of computational complexity at first, it is not entirely evident in general. 
In principle, changing the couplings could also lead to an increased amount of correlations offsetting the gain in the local Hilbert space dimension.
We will further elaborate on this issue in Section~\ref{Sec:Numerics} considering a specific example.

\subsection{Thermalized fTEDOPA} \label{Sec:thermalized_tedopa}
\subsubsection{First derivation}
\label{Sec:FTT_first_derivation}

Up to this point we recapitulated the thermofield-based approach~\cite{Vega2017} for fermions including non-zero chemical potential.
However, invoking further assumptions, the two parts of the Hamiltonian corresponding to the operators $e_\omega$ and $d_\omega$, respectively, can be mapped onto a single Hamiltonian term.
In particular, let us assume a vanishing chemical potential, $\mu = 0$, as well as a system-bath coupling that factorizes. 
The interaction Hamiltonian then reads
\begin{align}
	H_{\text{int}} &= L \; \int_{0}^{\wmax} \dw h(\omega) (f_\omega + f_\omega^\dagger).
\end{align}
Note that this implicitly assumes that $(L f_\omega^\dagger)^\dagger = L f_\omega$ due to the hermicity of $H_{\text{int}}$.
A specific example is the case of a single fermionic mode $f_{\text{sys}}$ coupled to the bath via $L := f_{\text{sys}} - f_{\text{sys}}^\dagger$ which is studied in more detail in~\cref{Sec:modified_rlm}.
Currently, this assumption is crucial and thermalized fTEDOPA does not work for other forms of the system-bath interaction such as a hopping interaction of the form $f_{\text{sys}}^\dagger f_\omega + f_\omega^\dagger f_{\text{sys}}$.
An extension to a wider class of interactions will be subject to future work.

Performing the thermofield transformation we end up with
\begin{align}
	\begin{split}
		H^\prime_{\text{int}} &= L \; \int_{0}^{\omega_{\max}} \dw h(\omega) u(\omega) \left( d_\omega + d_\omega^\dagger \right) \\
		&\hspace{10pt} + L \; \int_{0}^{\omega_{\max}} \dw h(\omega) v(\omega) \left(e_\omega^\dagger + e_\omega \right).
	\end{split}
	\label{Eq:thermalized_tedopa_thermofield_transformation}
\end{align}
Swapping $\omega \mapsto -\omega$ in the second term and using the fact that 
\begin{align}
    v(-\omega)^2 &= \frac{1}{\e^{-\beta \omega} + 1} = \frac{\e^{\beta\omega}}{\e^{\beta \omega} + 1} = u(\omega)^2 \label{Eq:mirror_coupling}
\end{align}
we obtain
\begin{align}
	\begin{split}
		H_{\text{int}} &= L \; \int_{0}^{\wmax} \dw h(\omega) u(\omega) \left( d_\omega + d_\omega^\dagger \right) \\
		&\hspace{10pt} + L \; \int_{-\wmax}^{0} \dw h(-\omega) v(-\omega) \left(e_{-\omega}^\dagger + e_{-\omega} \right)
	\end{split} \\
	&= L \int_{-\wmax}^{\wmax} \dw h(|\omega|)u(\omega) \left(\alpha_\omega^\dagger + \alpha_\omega\right).\label{Eq:thermalized_first_approach_interaction}
\end{align}
Here, we defined a new set of modes by
\begin{align}
    \alpha_\omega := 
    \begin{cases}
        d_\omega & \text{for } \omega \geq 0 \\
        e_{-\omega} & \text{for } \omega < 0.
    \end{cases}
    \label{Eq:thermalized_first_approach_new_modes}
\end{align}
in order to unify the two integrals in \cref{Eq:thermalized_first_approach_interaction}.
If $\mu \not= 0$, \cref{Eq:mirror_coupling} no longer holds and we cannot directly obtain the form of \cref{Eq:thermalized_first_approach_interaction}.

Similarly, the free Hamiltonian of the bath can be brought into the form
\begin{align}
    \tilde{H}^\prime_{\text{env}} &= \int_{0}^{\wmax} \dw \omega (d_\omega^\dagger d_\omega - e_\omega^\dagger e_\omega) \\
    &= \int_{0}^{\wmax} \dw \omega \; d_\omega^\dagger d_\omega + \int_{-\wmax}^0 \dw \omega \; e_{-\omega}^\dagger e_{-\omega} \\
    &= \int_{-\wmax}^{\wmax} \dw \omega\;\alpha_\omega^\dagger \alpha_\omega.
\end{align}
Note that the vacuum state w.r.t. the operators $d_\omega$ and $e_\omega$ is the same as the vacuum state w.r.t. $\alpha_\omega$.
Consequently, in this specific setting we can replace the two environments present in the thermofield formulation by a single environment comprising the same information.

We note two implications of this replacement.
Firstly, the thermalized fTEDOPA setting only requires the discretization of a single environment instead of two different environments.
This is advantageous because the numerical algorithms computing the discretization have access to all the information about the system-bath coupling at once. 
In contrast, the two baths obtained from the thermofield-based approach need to be discretized separately.
Thus, a discretization algorithm relies on a priori knowledge of the user to find optimally placed discrete modes.
We will further elucidate this effect in Section~\ref{Sec:Numerics}.

Secondly, consider the canonical situation in transport and quantum thermodynamical models where the system is initially coupled to two environments.
While we can easily deal with this scenario within thermalized fTEDOPA, the solution is much more involved in the thermofield-based framework.
This is due to the fact that the thermofield transformation would yield four environments in the first place.
Mapping these four environments onto a linear geometry is in principle possible, however, needs more sophisticated techniques.

\subsubsection{Second derivation}
\label{Sec:FTT_second_derivation}

We now present a different approach to devise thermalized fTEDOPA.
Consider again a Hamiltonian of the form
\begin{align}
	\begin{split}
		H &= H_{\text{sys}} + \int_{0}^{\wmax} \dw \omega f_\omega^\dagger f_\omega \\
		  &\hspace{10pt} + L \; \int_{0}^{\wmax} \dw h(\omega) (f_\omega + f_\omega^\dagger),
	\end{split}
\end{align}
where $H_{\text{sys}}$ denotes the free Hamiltonian of the system and $L$ its coupling operator.
By virtue of \cref{Eq:SpectralDensity} we can express the coupling function $h(\omega)$ in terms of the spectral density of the environment such that the coupling operator of the bath reads
\begin{align}
    A_{\text{env}} &:= h(\omega) \; (f_\omega + f_\omega^\dagger) = \sqrt{\frac{J(\omega)}{\pi}} \; (f_\omega + f_\omega^\dagger). \label{Eq:interaction_operator}
\end{align}
Given this Hamiltonian we want to simulate the reduced dynamics of the system under two assumptions.
Firstly, we demand the initial state to be in a product form between system and environment, i.e., $\rho(0) = \rho_{\text{sys}}(0) \otimes \rho_{\text{env}}(0)$.
Secondly, we only consider Gaussian initial states of the environment and in particular thermal states at inverse temperature $\beta$ and vanishing chemical potential $\mu = 0$ as defined in \cref{Eq:thermal_state_chemical_potential}, i.e., $\rho_{\text{env}}(0) = \rho_{\beta,0}$.
Under these assumptions, the reduced state is uniquely determined by the first and second moments of the bath interaction operator at all times.

This is due to the fact that starting from a factorized state and a factorized interaction Hamiltonian, the dynamics of the system are determined by the multi-time statistics of the bath interaction operator $A_{\text{env}}(t)$ (see Chapter 9 of Ref.~\cite{Breuer2016}).
Since $A_{\text{env}}(0)$ is linear in the creation and annihilation operators and the free Hamiltonian $H_{\text{env}}$ is Gaussian, $A_{\text{env}}(t)$ remains linear for all times. 
Hence, according to Wick's theorem, the multi-time statistics can be computed in terms of the two-times correlation function with respect to the bath state $\rho_{\beta,0}$ (see Ref.~\cite{Bach1994} and proof in Ref.~\cite{[Appendix A of ]Krumnow2017}).

Hence, any transformation that leaves the first and second moments invariant also preserves the dynamics of the system.
We exploit this fact and firstly show that the temperature dependence of the bath state can then be moved to the spectral density, i.e., the couplings $h(\omega)$ in the Hamiltonian.
Secondly, we prove that the first moments also stay the same under this transformation.

The second moments in terms of the interaction picture w.r.t. the free bath Hamiltonian are
\begin{align}
    C^{(2)}(t) &:= \int_{0}^{\wmax} \dw \langle A_{\text{env}}(\omega,t) A_{\text{env}}(\omega, 0) \rangle_{\rho_\beta}.
\end{align}
Substituting $A_{\text{env}}$ by \cref{Eq:interaction_operator} and $f_\omega(t) = \e^{\ii \omega t} \, f_\omega(0)$ we obtain
\begin{align}
	C^{(2)}(t) &= \int_{0}^{\wmax} \dw \langle A_{\text{env}}(\omega,t) A_{\text{env}}(\omega, 0) \rangle_{\rho_\beta} \;  \\
	\begin{split}
		&= \frac{1}{\pi} \int_{0}^{\wmax} \dw J(\omega) \; \left\langle (f_\omega^\dagger(t) + f_\omega(t)) \right. \\
		&\hspace{30pt} \left. \times (f_\omega^\dagger(0) + f_\omega(0)) \right\rangle_{\rho_\beta}
	\end{split} \\
	\begin{split}
		&= \frac{1}{\pi} \int_{0}^{\wmax} \dw J(\omega) \; \left[ \e^{\ii \omega t} \left\langle f_\omega^\dagger(0) f_\omega(0) \right\rangle_{\rho_\beta} \right. \\ 
		& \hspace{30pt} \left. + \e^{-\ii \omega t}  \left\langle f_\omega(0) f_\omega^\dagger(0) \right\rangle_{\rho_\beta} \right]
	\end{split} \\
	\begin{split}
		&= \frac{1}{\pi} \int_{0}^{\wmax} \dw J(\omega) \; \e^{\ii \omega t} \langle n_\beta(\omega) \rangle_{\rho_\beta} \; \\
		&\hspace{20pt}+ \int_{0}^{\wmax}  \dw J(\omega) \; \e^{-\ii \omega t} (1-\langle n_\beta(\omega) \rangle_{\rho_\beta}).
	\end{split} \label{eq:correlation_function_occupation}
\end{align}
Since free fermions obey the Fermi-Dirac statistics, we further have
\begin{align}
    \begin{split}
        \langle n_\beta(\omega) \rangle_{\rho_\beta} &= \bar{n}_\beta(\omega) = \frac{1}{\e^{\beta \omega} + 1}, \\
       1 - \langle n_\beta(\omega) \rangle_{\rho_\beta} &= 1 - \bar{n}_\beta(\omega) = \frac{\e^{\beta \omega}}{\e^{\beta \omega} + 1}.
    \end{split}
\end{align}
Substituting these expressions into the integrals we find
\begin{align}
	\begin{split}
		C^{(2)}(t) &= \frac{1}{\pi} \int_{0}^{\wmax} \dw J(\omega) \e^{\ii \omega t} \frac{1}{\e^{\beta \omega} + 1}  \\
		&\hspace{10pt} + \frac{1}{\pi} \int_{0}^{\wmax} \dw J(\omega) \e^{-\ii \omega t} \frac{\e^{\beta \omega}}{\e^{\beta \omega} + 1}.
	\end{split}
\end{align}
In order to unify these two integrals again we map $\omega \mapsto -\omega$ in the first integral such that 
\begin{align}
	\begin{split}
		C^{(2)}(t) &= \frac{1}{\pi} \int_{-\wmax}^{0} \dw J(-\omega) \e^{-\ii \omega t} \frac{1}{\e^{-\beta \omega} + 1}  \\
		&\hspace{10pt} + \frac{1}{\pi} \int_{0}^{\wmax} \dw J(\omega) \e^{-\ii \omega t} \frac{\e^{\beta \omega}}{\e^{\beta \omega} + 1}
	\end{split} \\
	&= \frac{1}{\pi} \int_{-\wmax}^{\wmax} \dw J(|\omega|) \frac{\e^{\beta \omega}}{\e^{\beta \omega} + 1}\; \e^{-\ii \omega t}  \\
	\begin{split}
    	&= \frac{1}{\pi} \int_{-\wmax}^{\wmax} \dw \frac{J(|\omega|)}{2} \\ &\hspace{15pt}\times\left[\tanh{\left(\frac{\beta \omega}{2}\right) + 1}\right] \; \e^{-\ii \omega t}.
	\end{split}
\end{align}
Finally, by defining the thermal spectral density as $J_\beta: [0,\wmax] \rightarrow [0,\infty)$,
\begin{align}
	    J_\beta(\omega) &:= \frac{J(|\omega|)}{2} \left[\tanh{\left(\frac{\beta \omega}{2}\right) + 1}\right], \label{Eq:thermalized_spectral_density_fermions}
\end{align}
we write
\begin{align}
    C^{(2)}(t) &= \frac{1}{\pi} \int_{-\wmax}^{\wmax} \dw J_\beta(\omega) \e^{-\ii \omega t}.
\end{align}
This is equivalent to the two-times correlation function we would obtain given a spectral density $J_\beta$ and starting from the vacuum state of the environment for any point in time $t$.
Furthermore, for the first moments of the interaction operator we find
\begin{align}
	\hspace{-5pt}C^{(1)}(t) &:= \int_{0}^{\wmax} \dw \langle A_{\text{env}}(\omega, t) \rangle_{\rho_\beta} \nonumber\\
	&= \frac{1}{\pi} \int_0^{\wmax} \dw J(\omega) \left( \langle f_\omega(t) \rangle_{\rho_\beta} + \langle f_\omega^\dagger(t) \rangle_{\rho_\beta} \right) \nonumber\\
	&= 0 \nonumber\\
	&= \frac{1}{\pi} \int_0^{\wmax} \dw J_\beta(\omega) \left( \langle f_\omega(t) \rangle_{|\Omega\rangle} + \langle f_\omega^\dagger(t) \rangle_{|\Omega\rangle} \right)
\end{align}
which completes our line of reasoning.

The two derivations are indeed equivalent. 
This can be seen by combining \cref{Eq:SpectralDensity,Eq:thermalized_first_approach_interaction,Eq:thermalized_spectral_density_fermions} 
which yields
\begin{align}
h(|\omega|)^2 u(\omega)^2 &= \frac{1}{\pi} J(|\omega|) \;\frac{\e^{\beta \omega}}{\e^{\beta \omega} + 1} \\
&= \frac{1}{\pi} \frac{J(|\omega|)}{2} \left[\tanh{\left(\frac{\beta \omega}{2}\right) + 1}\right] \\
&= \frac{1}{\pi} J_{\beta}(\omega).
\end{align}
Thus, we verify that the two coupling functions are identical.
This concludes our line of reasoning.

Finally, we want to comment once again on the necessity of a factorized interaction Hamiltonian in thermalized fTEDOPA.
The origin of this constraint can be most easily seen in the course of the first derivation described in Section~\ref{Sec:FTT_first_derivation}.
While the initial transformation to a thermofield Hamiltonian in \cref{Eq:thermalized_tedopa_thermofield_transformation} is still valid for non-factorizing interaction terms, the second step of merging the two auxilliary environments in \cref{Eq:thermalized_first_approach_new_modes} does not generalize straightforwardly.
Hence, in a scenario involving hopping interaction of the form $f_\text{sys}^\dagger f_\omega + f_\omega^\dagger f_\text{sys}$ we still have to resort to either the thermofield-based approach or the full density operator description.
For the sake of clarity, we provide an overview of the necessary assumptions on the model Hamiltonians for general, thermofield-based and thermalized fTEDOPA in \cref{Tb:comparison_model_assumptions}.
\begin{table}
	\caption{\textbf{Comparison of model assumptions}. 
	The different simulation methods studied in this work are presented in the three columns.
	The rows detail the form of Hamiltonian terms and values of chemical potential $\mu$ that can be handled by the different methods.
	We see that each of the methods admits an arbitrary form of the system Hamiltonian and a non-interacting fermionic Gaussian environment Hamiltonian. 
	While fTEDOPA and thermofield-based approaches allow for interaction Hamiltonians of arbitrary linear form and for non-zero chemical potentials, thermalized fTEDOPA only allows for interaction Hamiltonians that comprise a single tensor product of system and linear environment operator and for zero chemical potential.
	}
	\label{Tb:comparison_model_assumptions}
	{
	\renewcommand{\arraystretch}{1.5}
	\begin{tabular}{c|c|c|c}
		               & fTEDOPA   & Thermofield & Thermalized \\ \hline\hline
		$H_\text{sys}$ & arbitrary & arbitrary   & arbitrary \\ \hline
		$H_\text{env}$ & $\int \omega f_\omega^\dagger f_\omega \dw$  & $\int \omega f_\omega^\dagger f_\omega \dw$ & $\int \omega f_\omega^\dagger f_\omega \dw$\\ \hline
		$H_\text{int}$ & $\int (L f_\omega^\dagger + \text{h.c.})\text{d}\omega$ & $\int (L f_\omega^\dagger + \text{h.c.}) \dw$ & $L \int (f_\omega^\dagger + f_\omega) \text{d}\omega$\\ \hline
		$\mu$ & arbitrary & arbitrary & 0 
	\end{tabular}
	}
\end{table}

\section{Numerical examples}
\label{Sec:Numerics}

In this section we present various numerical examples involving model systems that highlight the capabilities of fTEDOPA and allow for benchmarking its accuracy against analytical methods wherever available.
Therefore, we will proceed in three different stages with increasing model complexity.
Additionally, we present a thorough case study of the three different simulation schemes for dealing with fermionic heat baths.
As we pointed out in Section~\ref{Sec:jordan_wigner_transformation} all Hamiltonians considered in this work transform to one-dimensional, nearest-neighbor, spin Hamiltonians which allows us to use TEBD to perform regular and imaginary time evolution throughout this section.

\subsection{Resonant level model}
\label{Sec:resonant_level_model}

First we demonstrate that fTEDOPA can simulate the well-studied resonant level model, for which an analytical solution is available.
The resonant level model describes a single fermionic mode, usually referred to as the impurity, coupled to a fermionic environment usually in the form of a sharply bounded, half-filled conduction band.
As it is conventional, we measure energy in terms of the half bandwidth such that all frequencies are confined to the interval $[-1,1]$.
Following the notation introduced in Section~\ref{Sec:Background} the Hamiltonian is split into three parts
\begin{align}
	H_{\text{rlm}} :=\,& H_{\text{sys}} + H_{\text{env}} + H_{\text{int}}, \label{Eq:rlm_hamiltonian}
\end{align}
where the individual parts are defined by
\begin{align}
	H_{\text{sys}} :=\,& E_{\text{imp}}\;d^\dagger d, \\
	H_{\text{env}} :=\,& \int_{-1}^{1} \dw \omega f_\omega^\dagger f_\omega, \label{Eq:rlm_env_hamiltonian}\\
	H_{\text{int}} :=\,& \int_{-1}^{1} \dw h(\omega) \left(d^\dagger f_\omega + f_\omega^\dagger d \right), \label{Eq:rlm_int_hamiltonian}
\end{align}
with $E_{\text{imp}}$ denoting the energy of the impurity.
Note that the impurity and the conduction band comprise the same species of fermions, i.e., $d$ and $f_\omega$ fulfill the fermionic CAR,
\begin{align}
\begin{split}
\{f_{\omega}, f_{\omega^\prime}\} &= \{d,d\} = \{f_{\omega}, d\} = \{f_{\omega}, d^{\dagger}\} = 0, \\
\{f_{\omega}, f_{\omega}^{\dagger}\} &= \delta(\omega - \omega^\prime), \\ 
\{d, d^\dagger\} &= 1.
\end{split}
\end{align}
Moreover, the system-bath coupling $h(\omega)$ is directly related to the spectral density as shown in \cref{Eq:sd_coupling_linear}.
We choose the spectral density to be frequency independent, i.e.,
\begin{equation}
	J: [-1,1] \rightarrow \R, \; \omega \mapsto \Gamma = \text{const}, 
\end{equation}
which results in a system-bath coupling of the form $h(\omega) = \sqrt{\frac{\Gamma}{\pi}}$.

We are interested in the relaxation of the occupation of the impurity after a sudden quench in its energy.
More precisely, we prepare the system initially in the ground state of $H_{\text{rlm}}$ with $E_{\text{imp}}(t < 0) = 0$ and switch to $E_{\text{imp}}(t\geq0) \not= 0$ at $t = 0$.
Since $E_{\text{imp}}(t < 0)$ is equal to the Fermi energy of the conduction band we expect the initial occupation to be $\bar{n}_{\text{imp}} := \langle d^\dagger d \rangle = 0.5$.
After the quench the occupation is expected to equilibrate to a higher (lower) value depending on whether $E_{\text{imp}}(t \geq0)$ is lower (higher) than the Fermi energy.

As discussed in Section~\ref{Sec:state_preparation} the ground state is no longer the vacuum state as soon as the system is subject to a non-vanishing chemical potential. 
Hence, for a half-filled conduction band, i.e., a chemical potential which is equal to half the bandwidth, the initial state has to be prepared numerically.
This can be done in various ways such as imaginary time evolution or variational DMRG.
Both algorithms are explained in great detail in any comprehensive review on tensor network techniques, e.g., Ref.~\cite{Schollwoeck2011}.
Within this work we will always prepare ground states using variational DMRG.

In order to verify the results from fTEDOPA we fix a proper reference solution.
While there exists an analytical solution of the resonant level model using the Keldysh formalism, these solutions typically rely on a wide-band approximation of the conduction band~\cite{Anders2006}.
The wide-band approximation is based on the assumption that the strength of the system-bath interaction is much weaker than the bandwidth of the conduction band, i.e., $\Gamma \ll 1$ in our setting.
Taking this solution as a reference would make it difficult to distinguish between any error introduced by the violation of the wide band limit and the errors arising within fTEDOPA.
However, since the model comprises only one species of fermions there is an alternative way to obtain a reference solution.
More precisely, in this case the Hamiltonian in \cref{Eq:rlm_hamiltonian} can be interpreted as a bilinear form of the fermionic creation and annihilation operators.
Applying a suitable discretization technique one is able to construct an approximating, discrete bilinear form.
It is well known that this kind of discrete systems can then in turn be analytically and numerically treated by so called exact diagonalization (ED)~\cite{[Appendix A of ]Serafini2017}.
Numerically, this approach involves linear algebra operations on a matrix whose size scales only linearly in the number of discrete modes.
Thus, in contrast to the wide-band approximation, here it is easier to control the quality of the approximation by increasing the number of sampling points in the conduction band.

Consequently, we generate our reference solution in two steps.
First, we discretize the continuous Hamiltonian in \cref{Eq:rlm_hamiltonian}.
Here, it is important to choose a discretization technique that is different from fTEDOPA in order to obtain a meaningful benchmark.
Hence, we choose a linear discretization of the energy domain $[-1,1]$ with a high number of modes $N_{\text{ED}}$ which yields a discrete and finite Hamiltonian.
Second, we use standard ED to determine the initial state and the dynamics of the impurity after the quench.

\begin{figure}[t]
	\includegraphics[width=0.98\columnwidth]{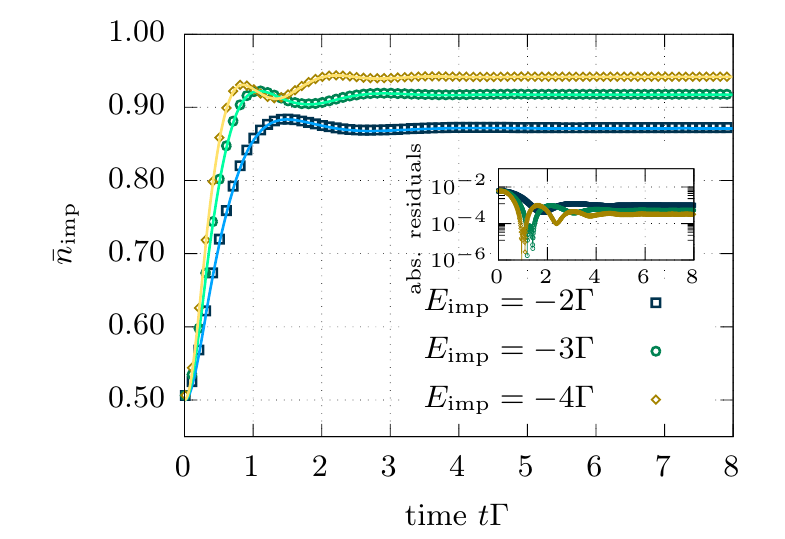}
	\caption{\textbf{Simulation of an energy quench in the resonant level model:}
	Occupation of the impurity as a function of time after a quench of the energy from $E_{\text{imp}}(t<0) = 0.0$ to $E_{\text{imp}}(t\geq0) = -2\Gamma,\,-3\Gamma,\,-4\Gamma$.
	Here, the conduction band is described by the constant spectral density $J: [-1,1] \rightarrow \R,\, \omega \mapsto \Gamma = 0.1$.
	The results were obtained using a chain of $N=128$ sites, a relative truncation error of $\epsilon_{\text{trunc}}^{\text{te}}=10^{-8}$ and a maximal truncation rank of $\xi_{\text{trunc}}^{\text{te}}=128$ in the compression.
	The time step width is set to $\text{d}t=0.1$ and a Trotter expansion of $4^\text{th}$ order is used.
	The initial ground state is computed with 10 DMRG sweeps, a relative truncation error of $\epsilon_{\text{trunc}}^{\text{gs}}=10^{-8}$ and a maximal truncation rank of $\xi_{\text{trunc}}^{\text{gs}}=64$.
	As a reference we plot with solid lines the results obtained from exact diagonalization using a chain of $N_{\text{ED}}=500$ sites.
	The inset depicts the absolute residuals $|\bar{n}(t) - \bar{n}_{\text{ED}}(t)|$ between the ED and the fTEDOPA solution.
	See \cref{Sec:resonant_level_model} for more details.}
	\label{fig:occupation_rlm}
\end{figure}

\cref{fig:occupation_rlm} shows a comparison of the results obtained via fTEDOPA and ED.
For all three different choices of $E_{\text{imp}}(t\geq0)$ we observe good agreement of the impurity's occupation for moderate chain length and bond dimensions.
This observation is confirmed by the absolute residuals $|\bar{n}(t) - \bar{n}_{\text{ED}}(t)|$ depicted in the inset of \cref{fig:occupation_rlm}.
Furthermore, the residuals reveal that we start off with a relatively high error which then drops to a lower level as simulation time increases.
This effect is due to the imperfect initial ground state we start off with.
In principle, this could be improved by allowing for larger computational resources in the preparation of the initial state.

\subsection{Modified resonant level model}
\label{Sec:modified_rlm}

Extending the previous example, in this section, we examine an impurity that is subject to an environment at non-zero temperature.
In particular, we present a detailed study of the performance of standard MPO, thermofield and thermalized fTEDOPA evolution.
Here, we focus on comparing the growth of correlations, the memory costs and the achieved accuracy.
The results obtained may serve as a useful guideline for other applications.

In order to obtain a model that is accessible to thermalized fTEDOPA we consider a modified version of the resonant level model Hamiltonian \cref{Eq:rlm_hamiltonian}.
Firstly, we no longer assume a half-filled but an empty conduction band at zero temperature.
Hence, by shifting the chemical potential as well as the zero of energy to the bottom of the band, the free Hamiltonian of the conduction band reads
\begin{align}
    H_{\text{env}} &:= \int_{0}^{2} \dw \omega f_\omega^\dagger f_\omega.
\end{align}
Secondly, we choose the interaction Hamiltonian to be in the factorized form
\begin{align}
	H_{\text{int}} &:= (d^\dagger - d) \int_{0}^{2} \dw h(\omega) (f_\omega + f_\omega^\dagger).
\end{align}
Hence, the total Hamiltonian can be summarized as
\begin{align}
	H_{\text{mrlm}} &:= H_{\text{sys}} + H_{\text{env}} + H_{\text{int}}.
\end{align}
where $H_{\text{sys}} := E_{\text{imp}} d^\dagger d$ denotes the Hamiltonian of the impurity with energy $E_{\text{imp}}$.
The spectral density in turn is again chosen to be constant, i.e., 
\begin{align}
    J: [0,2] \rightarrow \R, \; \omega \mapsto \Gamma = \text{const}.
\end{align}
which leads to a constant system-bath coupling $h(\omega) = \sqrt{\frac{\Gamma}{\pi}}$ via \cref{Eq:sd_coupling_linear}.

Given the Hamiltonian we prepare the whole system in the initial state $\rho(0) = |1\rangle \langle1|_{\text{sys}} \otimes \rho_{\text{env}}(\beta)$ where $\rho_{\text{env}}(\beta)$ is a thermal state of the free environment at inverse temperature $\beta$.
At $t=0$ we then turn on the system-environment coupling and let the system equilibrate.

We study the following question.
Given a fixed number of sites, a fixed amount of memory resources and a maximally tolerable absolute error.
Which method allows for the largest possible simulation time (as defined by the system time at which the maximally tolerable absolute error is exceeded) while maintaining the lowest memory complexity?
It is important to emphasize that even though real processing time costs, i.e. CPU time, is very important in applications, it is not a good figure of merit here.
This is due to following reasons.
Firstly, the CPU time strongly depends on the implementation.
Secondly, even for the same implementation using iterative linear algebra solvers can lead to huge fluctuations in the computational time depending on the exact parameters of the Hamiltonian.

\begin{figure*}[htbp]
	\includegraphics[width=1.0\textwidth]{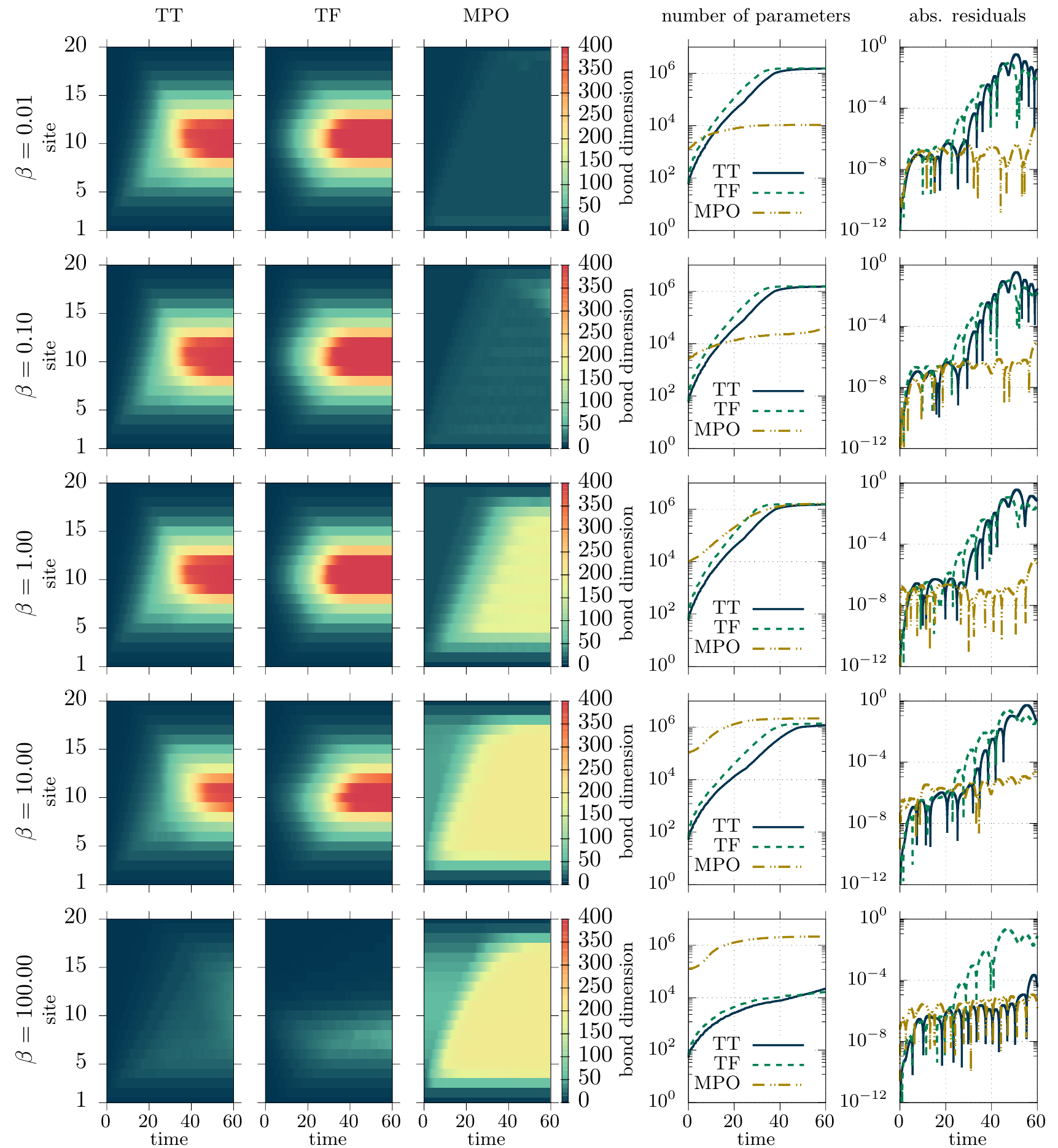}
	\caption{\textbf{Comparison of the performance of the different thermal evolution schemes for the modified resonant level model:}
	Evolution of the MPS/MPO bond dimensions of the quantum state evolved under the thermalized (first column), thermofield-based (second column) and MPO (third column) simulation scheme for different inverse temperatures $\beta$ ($\hbar = k_B = 1$).
	As a precise measure of memory complexity the total number of parameters of the MPS/MPO as a function of time is plotted in the fourth column.
	The fifth column depicts the absolute residuals $|\bar{n}(t) - \bar{n}_{\text{ED}}(t)|$ of the impurity's occupation w.r.t. the results obtained by ED with $N_\text{ED} = 300$ sites.
	The numerical results are calculated using a constant spectral density $J: [0,2] \rightarrow \R,\, \omega \mapsto \Gamma = 0.4$ and an impurity energy of $E_\text{imp} = 0.3$.
	In all three cases we fix the total number of sites to $N=20$, the relative truncation error to $\epsilon_{\text{trunc}}^\text{te}=10^{-8}$ and the maximal truncation rank of $\xi_{\text{trunc}}^{\text{te}}=400$ in case of the MPS routines and $\xi_{\text{trunc}}^{\text{te}} = 200$ in case of the MPO routine.
	Moreover, the time step width is set to $\text{d}t = 0.01$ and a Trotter expansion of $4^\text{th}$ order is used.
	In case of the MPO scheme the initial thermal state of the environment is computed via imaginary time evolution and exactly the same settings as for the real time evolution.
	\label{fig:fermionic_toymodel}}
\end{figure*}

\cref{fig:fermionic_toymodel} provides a thorough comparison of the results of the three simulation schemes, thermalized fTEDOPA (TT), thermofield (TF)~\cite{Vega2015a} and standard matrix product operator evolution (MPO) for different inverse temperatures $\beta$.
The figure is split into three main parts.
The first part depicts the growth of bond dimensions in time for each of the three methods.
These heatmaps not only allow us to identify the parts of the system that increase the computational complexity of the tensor-network representations but also enable us to study how correlations built up in the system depending on the different algorithms.
Since the bond dimensions itself do not directly tell us the memory complexity of the individual state representations due to the different local dimensions, the second part shows the actual number of parameters to be stored at each time.
In the last part of the figure as a measure of accuracy we plot the absolute residuals $|\bar{n}(t) - \bar{n}_{\text{ED}}(t)|$ of the occupation of the impurity.

Based on these three parts we summarize the following findings.
For small $\beta$, i.e., high temperatures $T = 1/\beta$, the correlations, as quantified by the bond dimensions, grow very rapidly in both TT and TF, while for MPO there are hardly any correlations visible.
This fact has important implication for the memory complexity.
In fact, TT and TF start off with a memory complexity roughly two orders of magnitude below MPO but end up with a complexity more than two orders of magnitude higher than MPO.
Moreover, the residuals in the occupation of the impurity start rising for TT after around half the time it takes for MPO.
For TF this rise happens even slightly earlier in time.
As $\beta$ increases, i.e., $T$ decreases, the growth of correlations becomes more prominent for MPO and less prominent for TT and TF.
This effect is also visible in the second part of the figure where the memory complexity of MPO increases about two orders of magnitude going from $\beta = 0.01$ to $\beta = 100.0$.
However, the accuracy remains unchanged for all three methods up to $\beta = 10.0$.

For large $\beta$, i.e. small temperatures $T$, the correlations behave differently for TF, TT and MPO as compared to the case of small $\beta$.
More precisely, the states involved in TT and TF show little correlations whereas the bond dimensions quickly saturate for MPO.
The same holds for the memory complexity where MPO suffers a memory complexity two orders of magnitude higher than TT and TF throughout almost the whole time evolution.
Furthermore, we see an interesting behavior of the accuracies in this temperature regime.
While TT and MPO perform very well throughout the whole time evolution of 60 time units, in TF the errors start rising after 20 units. 
This can be explained as follows.

Given a fixed number of bath sites $2N$ and without prior knowledge about the thermal spectral density, the discrete modes in TF are distributed equally w.r.t. the left and the right environment.
That means that each bath is straightforwardly sampled with the same number of modes $N$.
For higher values of $\beta$, however, the spectral density corresponding two the second term in \cref{eq:btraf_interaction_hamiltonian} is strongly suppressed due to the Fermi-Dirac distribution.
Hence, one of the environments couples severely weaker to the system than the other.
An equal distribution therefore leads to the over-sampling of an unimportant portion of the overall spectral density.
This problem is circumvented in TT and MPO since we only have to deal with one spectral density and the available modes are dynamically distributed among the relevant parts of the spectral density. 
Nevertheless, so far, TT only addresses interaction Hamiltonians that are in a tensor product form, which excludes a variety of important physically relevant cases for fermionic systems.
In these situations MPO and TF are the only options.

In summary, our simulations can provide guidelines about which simulation method to use out of TT, TF or MPO.
Surprisingly, TT and TF, two simulations schemes based on MPSs are not always favorable as compared to the MPO method but the method of choice should strongly depend on the specific setting.
We observe that MPO is more suitable in case of low $\beta$ while for high $\beta$ TT or TF seem to be better suited. 
Moreover, whenever it is possible to use TT (i.e., if the interaction Hamiltonian is in a tensor product form) it is preferable to TF.
Last but not least, we want to point out that series expansions of spin-systems as well as strongly correlated electrons have been studied extensively in the past for both low and the high temperature~\cite{Oitmaa2006}. 
In their respective domains of validity, these methods could potentially outperform any of the given tensor-based techniques for certain applications.
However, this is still an open question and must be subject of future work.

\subsection{Quantum dot}

As a third example, we study a Hamiltonian used to model the decoherence of a quantum dot.
The model comprises a two-level system (TLS) coupled to a fermionic impurity which itself is subject to a half-filled conduction band.
This kind of setting is for example used to study decoherence effects in Josephson junctions~\cite{Girshin2005,Paladino2002a,Abel2008}.

Following Section~\ref{Sec:Background} we split the Hamiltonian of the model into three parts,
\begin{align}
	H_{\text{qd}} &:= H_\text{sys} + H_{\text{env}} + H_{\text{int}}.
\end{align}
Parameterizing the energy separation of the TLS by $\Delta$, the energy of the impurity by $E_{\text{imp}}$ and the coupling strength between TLS and impurity by $v$, we define
\begin{align}
	H_\text{sys} &:= \frac{\Delta}{2} \sigma_z + \frac{v}{2} \sigma_z \; d^\dagger d + E_{\text{imp}} d^\dagger d.
\end{align}
Moreover, the environment and interaction Hamiltonian are defined analogously to the resonant level model by \cref{Eq:rlm_env_hamiltonian} and \cref{Eq:rlm_int_hamiltonian}, respectively.
The spectral density is chosen to be
\begin{align}
	J: [-1,1] \rightarrow \R, \omega \mapsto \Gamma = \text{const},
\end{align}
which again leads to a constant system-bath coupling of the form $h(\omega) = \sqrt{\frac{\Gamma}{\pi}}$ via \cref{Eq:sd_coupling_linear}.

We study the decoherence effects on the TLS by preparing the total system initially in a factorized state $\rho(0) = |+\rangle\langle+|_{\text{tls}} \otimes \rho_{\text{elec},0}$ between the TLS and the electronic part of the Hamiltonian, namely the impurity and the conduction band.
Here, the TLS is in a superposition state
\begin{align}
	|+\rangle_{\text{tls}} := \frac{1}{\sqrt{2}} (|0\rangle + |1\rangle), \label{Eq:plus_state}
\end{align}
and $\rho_{\text{elec},0}$ denotes the ground state of the impurity coupled to the half-filled conduction band.
As done for the case of the resonant level model, we compute the ground state numerically using variational DMRG.
Subsequently, we turn on the interaction $v$ between the TLS and the impurity, and let the system evolve in time.
During this evolution we monitor the coherence of the TLS in terms of the quantity
\begin{align}
	C(t) &:= \left|\frac{\rho_{\text{tls},01}(t)}{\rho_{\text{tls},01}(0)} \right| = 2 |\rho_{\text{tls},01}(t)|.
	\label{Eq:coherence_tls}
\end{align}
Studying this time dependence is crucial in the applications of quantum dots to quantum technologies as qubits or as single photon sources~\cite{Wolf2001a}.

\begin{figure}[t]
	\includegraphics[width=\columnwidth]{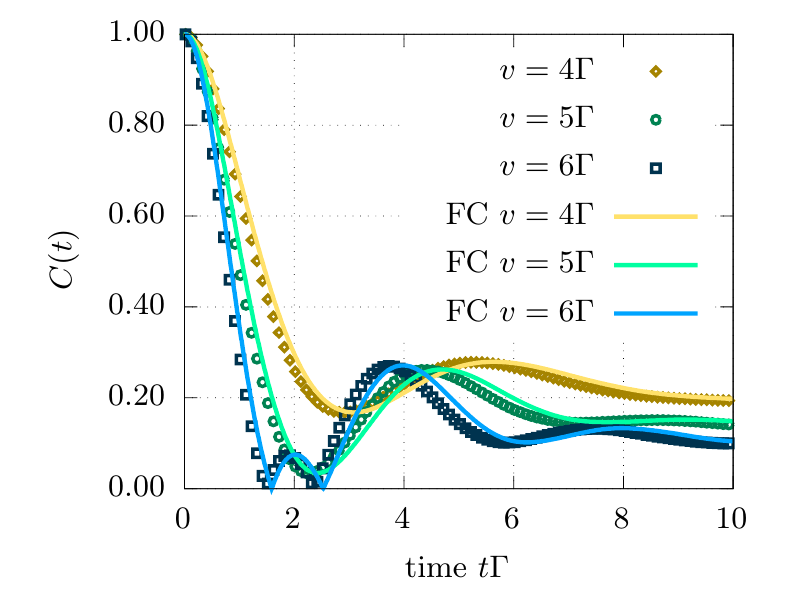}
	\caption{\textbf{Simulation of quantum dot coupled to fermionic reservoir:}
	Coherence of the TLS, defined in \cref{Eq:coherence_tls}, as a function of time and different coupling strengths $v$ between the TLS and the impurity.
	The initial state is chosen to be $\rho(0) = |+\rangle\langle+|_{\text{tls}} \otimes \rho_{\text{elec},0}$ where $|+\rangle_{\text{tls}}$ is defined in \cref{Eq:plus_state} and $\rho_{\text{elec},0}$ denotes the ground state of the impurity coupled to the conduction band.
	The conduction band is described by a constant spectral density $J: [-1,1] \rightarrow \R,\, \omega \mapsto \Gamma = 0.1$, the energy of the impurity is set to the Fermi energy, i.e., $E_{\text{imp}} = 0.0$, and the energy splitting of the TLS is chosen to be $\Delta=0.2$.
	The numerical results were obtained using a chain of length $N=400$, a relative truncation error of $\epsilon_{\text{trunc}}^{\text{te}}=10^{-8}$ and a maximal truncation rank of $\xi_{\text{trunc}}^{\text{te}}=192$ in the compression.
	The time step width is set to $\text{d}t=0.1$ and a Trotter expansion of $4^\text{th}$ order is used.
	The initial ground state is obtained via 10 DMRG sweeps with a relative truncation error of $\epsilon_{\text{trunc}}^{\text{gs}}=10^{-8}$ and a maximal truncation rank of $\xi_{\text{trunc}}^{\text{gs}}=64$.
	We compare the fTEDOPA results to results based on full-counting (FC) statistics for $N_{\text{FC}} = 1000$ modes following Ref.~\cite{Abel2008}.
	\label{fig:quantum_dot}}
\end{figure}

We compare our solution against the one obtained under the assumption of a wide-band limit ($\Gamma \ll 1$). 
Following the full-counting (FC) statistics approach of Ref.~\cite{Abel2008}, we compute the time evolution of $C$, which is depicted in \cref{fig:quantum_dot} for fixed $\Delta$, $E_{\text{imp}}$ and $\Gamma$ and various $v$.
This reference solution is expected to provide good results for small values of $\Gamma$ but is expected to be inaccurate for larger values. 
The fTEDOPA solutions match the reference solution well in terms of the shape, but we see a slight difference in the phase.
In order to verify the fTEDOPA results we checked convergence by varying the chain length, the maximal bond dimension, the time step width as well as the number of DMRG sweeps to generate the ground state of the environment.
Furthermore, our results show good agreement in shape and phase with the results obtained in a detailed study of the same model using linear and logarithmic discretization techniques~\cite{Schroeder2012}.
Hence, we suspect that the shift in phase is indeed due to the violation of the wide-band assumption and thus fTEDOPA is more accurate than the well-established full-counting statistics approach.

\begin{figure*}[t]
	\includegraphics[width=0.98\textwidth]{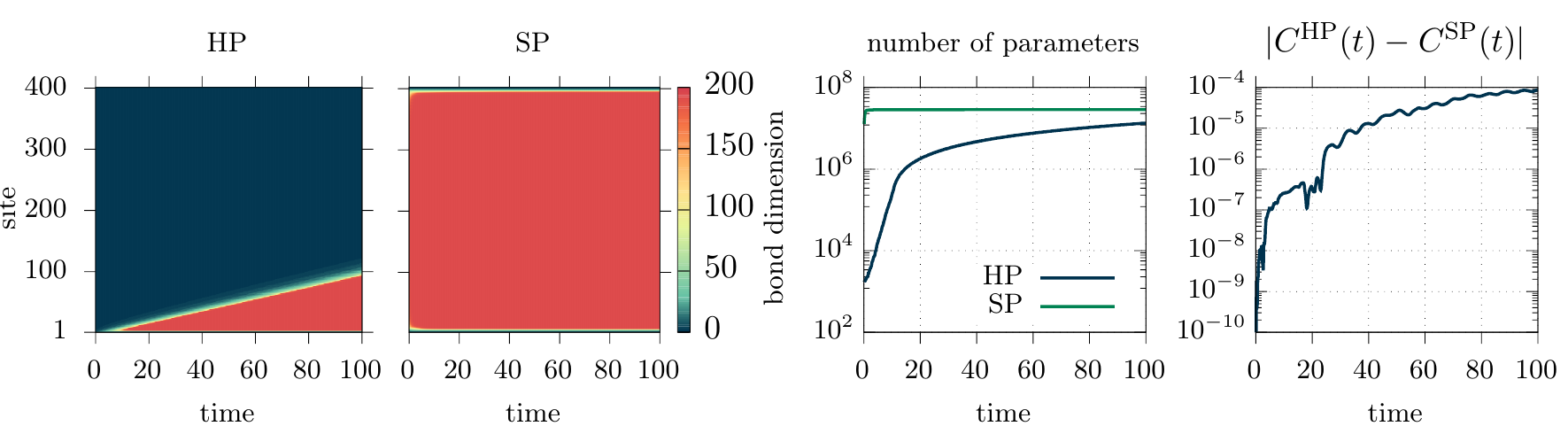}
	\caption{\textbf{Memory complexity in Heisenberg and Schr\"odinger picture for quantum dot coupled to fermionic environment.}
	The first two panels depict the growth of bond dimensions in the Heisenberg picture (HP) and the Schr\"odinger picture (SP), respectively.
	The third panel compares the actual number of parameters to be stored.
	In the last panel we show the absolute residuals between Heisenberg and Schr\"odinger solution, i.e., $|C^{\text{HP}}(t) - C^{\text{SP}}(t)|$.
	For the comparison we used the same simulation parameters as described in \cref{fig:quantum_dot} with $v = 6\Gamma$.}
	\label{fig:quantum_dot_heisenberg}
\end{figure*}

As we pointed out in Section~\ref{Sec:state_preparation}, in the presence of a chemical potential, the ground state is expected to be highly correlated and thus needs large bond dimension for obtaining an accurate MPS representation.
In general the bond dimensions further increase over time rendering simulations very costly.
However, this reasoning implicitly assumes that the time evolution is carried out in the Schr\"odinger picture, i.e. the state itself is propagated in time.

In some cases, the issues arising here might be circumvented by switching to the Heisenberg picture where the state itself is time-independent and rather the observable of interest is evolved in time~\cite{Hartmann2009, Clark2010}.
\cref{fig:quantum_dot_heisenberg} shows that, when simulating the coherence of the quantum dot as a function of time, the performance of TEDOPA is drastically improved in the Heisenberg picture.
This may as well translate to a much wider set of fermionic open quantum systems which will be explored and characterized in future work.

\subsection{Dimeric transport problem}

As a final example of the versatility of fTEDOPA, we simulate a paradigmatic model in quantum transport and thermodynamics namely that of two coupled quantum dots~\cite{Wiel2002a}.
In particular, we consider a dimeric system initially disjoint from two separate heat baths which then evolves to a steady state.
The Hamiltonian of the model can be split into five terms,
\begin{align}
    H_{\text{dim}} &:= \sum_{\alpha = \text{L},\text{R}} (H_{\text{env},\alpha} + H_{\text{int}, \alpha}) + H_{\text{sys}}. \label{Eq:HDimer}
\end{align}
Here, the system Hamiltonian is 
\begin{align}
	\begin{split}
	    H_{\text{sys}} &:=  \sum_{\gamma = \text{L},\text{R}} h_\gamma d_\gamma^\dagger d_\gamma \\
	    &\hspace{15pt} - \frac{g}{2} (d_{\text{L}}^\dagger d_{\text{R}} + d_{\text{R}}^\dagger d_{\text{L}}) + U n_{\text{L}} n_{\text{R}}
	\end{split}
\end{align}
where $g$ and $U$ denote the intra-system coupling and the Coulomb interaction strength, respectively.
The energy level structure of the system is parameterized by the mean energy $h = \frac{1}{2}(h_{\text{L}} + h_{\text{R}})$ as well as the detuning $\delta = h_{\text{L}} - h_{\text{R}}$.
Assuming the two reservoirs share the same conduction band characteristics, we again measure energy in terms of the half bandwidth.
Thus, the Hamiltonians of the fermionic reservoirs read
\begin{align}
    H_{\text{env},\alpha} &:= \int_{0}^{2} \dw \omega f_{\alpha,\omega}^\dagger f_{\alpha,\omega}, \\
	H_{\text{int},\alpha} &:= \int_{0}^{2} \dw h(\omega) (d_\alpha^\dagger f_{\alpha,\omega} + f_{\alpha,\omega}^\dagger d_\alpha)
\end{align}
for $\alpha = L,R$.
Furthermore, we consider the so-called Newns' spectral density~\cite{Newns1969}
\begin{align}
	J : [0, 2] \rightarrow \R, \; \omega \mapsto \frac{\Gamma}{2} \sqrt{1 - (\omega - 1)^2}.
	\label{Eq:Newns_spectral_density}
\end{align}
which, after exploiting~\cref{Eq:sd_coupling_linear}, yields a system-bath coupling $h(\omega)$ of the form 
\begin{align}
	h: [0, 2] \rightarrow \R, \; \omega \mapsto \left(\frac{\Gamma}{2\pi}\right)^{\frac{1}{2}} \left( 1 - (\omega - 1)^2\right)^{\frac{1}{4}}.
\end{align}

\begin{figure*}[htbp]
	\includegraphics[width=0.98\textwidth]{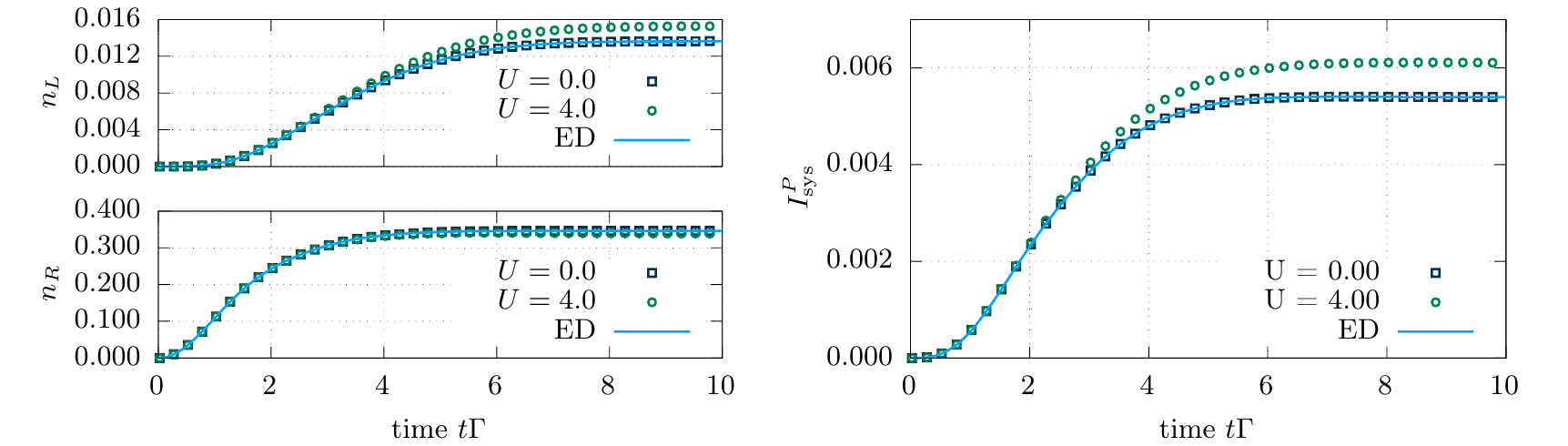}
	\caption{\textbf{Evolution of a dimeric system subject to two distinct fermionic reservoirs to its steady-state:}
	The two left panels show the occupation of the left and right site of the system, respectively, while the right panel depicts the intra-system particle current $I_{\text{sys}}^P$ defined in \cref{Eq:intra_sys_current}.
	The initial state is chosen as $\rho(0) = |\textbf{0} \rangle \langle \textbf{0}|_{\text{L}} \otimes |0,0\rangle \langle 0,0|_{\text{sys}} \otimes \rho_{\text{R},\beta}$ where $|\textbf{0}\rangle_{\text{L}}$ and $|0,0\rangle_{\text{sys}}$ denote the vacuum state of the left reservoir and the system, respectively, and $\rho_{\text{R},\beta}$ denotes the thermal state of the right bath at inverse temperature $\beta = 1.0$ and vanishing chemical potential, $\mu = 0$.
	Both reservoirs are subject to the Newns' spectral density defined in \cref{Eq:Newns_spectral_density} with $\Gamma = 0.5$ and comprise $N_\alpha=10$ ($\alpha = L,R$) sites each.
	The thermal state of the right bath is prepared via imaginary time evolution using a time step width $\text{d}t=0.1$, a Trotter expansion of $4^\text{th}$ order, a relative truncation error of $\epsilon_{\text{trunc}}^{\text{ite}}=10^{-8}$ and a maximal truncation rank of $\xi_{\text{trunc}}^{\text{ite}}=128$.
	The Coulomb interaction strength is varied between $U = 0$ and $U = 4.0$.
	For the subsequent real time evolution we choose the mean energy of the system to be $h = 0.6$, the detuning to be $\delta = 0.01$ and the intra-system coupling strength to be $g = 0.1$.
	Furthermore, we use a Trotter decomposition of $4^\text{th}$ order, a time step width of $\text{d}t=0.05$, a relative truncation error of $\epsilon_{\text{trunc}}^{\text{te}}=10^{-8}$ and a maximal truncation rank of $\xi_{\text{trunc}}^{\text{te}}=256$.
	In case of a vanishing Coulomb interaction we also depict the results of exact diagonalization (ED) with $N_{\text{ED}} = 200$ with solid lines.}
	\label{fig:dimer}
\end{figure*}

For the numerical example at hand we initially prepare the total system in the product state 
\begin{equation}
	\rho(0) = \rho_{\text{L},0} \otimes \rho_{\text{sys},0} \otimes \rho_{\text{R},\beta},
\end{equation}
where $\rho_{\text{L},0}$ and $\rho_{\text{sys},0}$ denote the ground state of the left reservoir and the system, and $\rho_{\text{R},\beta}$ denotes the thermal state of the right reservoir at inverse temperature $\beta$.
Note that in principle all components could be subject to a chemical potential $\mu$.
However, within the scope of this example we set $\mu = 0$.
Thus, ground states effectively reduce to vacuum states and need no special preparation.
After switching on the system-bath couplings at $t = 0$ we then observe the relaxation of the global system towards an approximately steady state. 
We monitor the occupation of the left and right system site as well as the intra-system particle current,
\begin{align}
	I_{\text{sys}}^P := \frac{g}{2\ii} (d_{\text{L}}^\dagger d_{\text{R}} - d_{\text{R}}^\dagger d_{\text{L}}), \label{Eq:intra_sys_current}
\end{align}
over time.
Note that the fermionic degrees of freedom of the system and the reservoirs are again indistinguishable and thus obey the fermionic CAR.
Hence, for vanishing Coulomb interaction strength, i.e., $U = 0$, the Hamiltonian can again be interpreted as a bilinear form of the fermionic creation and annihilation operators and is thus accessible to exact diagonalization (ED) techniques.
The results obtained in this way serves as a reference for the fTEDOPA results.

Previous work~\cite{Meir1993a,Cronenwett1998a} has studied the case of $g = 0$, $U \ne 0$ in \cref{Eq:HDimer} and also~\cite{Brask2015a,Mitchison2018,Wang2019a} on $U = 0$, $g \ne 0$.
Moreover, similar models with $U,g\ne0$ have been considered~\cite{Schwarz2018}.
However, here the continuous environments are discretized by a combination of linear and logarithmic discretization.
Thus, fTEDOPA establishes a new way of simulating open quantum system considered in the field of quantum transport and quantum thermodynamics.

In \cref{fig:dimer} we depict the evolution of the system to its steady-state.
In case of vanishing Coulomb interaction the fTEDOPA results show good agreement with the results obtained via ED.
As soon as we increase the Coulomb interaction strength the steady-state occupations and the intra-system current get shifted.
While the occupation on the left site of the system and the intra-system current rise, the occupation of the right site of the system lowers.

Finally, we want to point out that --- as in the case of the quantum dot --- switching to the Heisenberg picture might again improve the performance of fTEDOPA.
In fact, for vanishing Coulomb potential, i.e., for a Hamiltonian that is completely Gaussian, it has been shown~\cite{Hartmann2009, Clark2010} that the time-evolved fermionic ladder operators obey an exact MPO representation with constant bond dimension.
In particular, after mapping the continuous setting onto a chain of total length $2N+2$, any fermionic operator $g_m(t)$, $m \in \{1,\dots,2N+2\}$, can be written in the form
\begin{align}
g_m(t) &:= \sum_{\ell = 1}^{2N+2} \left( \alpha_{m,\ell}(t) g_\ell(0) + \beta_{m,\ell}(t) g_\ell^\dagger(0) \right)
\label{Eq:time_evolution_fermionic_operators}
\end{align}
where $\alpha_{m,\ell}$ and $\beta_{m,\ell}$ are complex-valued functions in time.
Applying the Jordan-Wigner transformation defined in \cref{Eq:jordan_wigner_transformation} to \cref{Eq:time_evolution_fermionic_operators} yields
\begin{align}
\sigma^+_m(t) &= \sum_{\ell = 1}^{2N+2} \left( \bigotimes_{k = 1}^{\ell} \sigma_k^{z} \right) \left( \alpha_{m,\ell}(t) \sigma_\ell^{-} + \beta_{m,\ell}(t) \sigma_\ell^{+} \right).
\end{align}
This string of spin operators can in turn be expressed as an MPO with bond dimension equal to 2.
Adopting the notation of \cref{Eq:mpo_representation} the matrices of the MPO are defined by
\begin{align}
\begin{split}
[A_{j_1}]_{j_1} &:= 
\begin{bmatrix}
G_{m,1}(t) & \sigma_1^{z}
\end{bmatrix}, \\
[A_{j_n, j_n^\prime}]_{j_n, j_n^\prime} &:= 
\begin{bmatrix}
	\openone & 0 \\
	G_{m,\ell}(t) & \sigma_{\ell}^{z}
\end{bmatrix}, \\
[A_{j_{2N+2}}]_{j_{2N+2}} &:= 
\begin{bmatrix}
\openone \\
G_{m,2N+2}(t)
\end{bmatrix},
\end{split}
\end{align}
where $G_{m,\ell} := \alpha_{m,\ell}(t) \sigma_\ell^{-} + \beta_{m,\ell}(t) \sigma_\ell^{+}$, $\ell = 1,\dots,2N+2$.
Natural observables in the transport setting such as the occupation number operator or the intra-system particle current, \cref{Eq:intra_sys_current}, are then obtained by multiplying and adding the MPOs of the fermionic operators.
While the multiplication of two MPOs leads to an MPO with bond dimension equal to the product of the initial bond dimensions, the addition results in an MPO with bond dimension equal to the sum of the initial bond dimensions.
Hence, the number operator and the intra-system particle current admit an exact MPO representation with a bond dimension of at most 4 and 8, respectively.

Unfortunately, as soon as the Coulomb interaction is non-vanishing, the rigorous results presented in the previous paragraph break down.
However, for small Coulomb interaction strength, the Hamiltonian is only slightly perturbed from a Gaussian Hamiltonian which could lead to MPO representations of the fermionic operators whose bond dimension grow very slowly in time.
An in-depth study of the relative merits of simulations in the Schr\"odinger and the Heisenberg picture is warranted but beyond the scope of this work.

\section{Discussion}
\label{Sec:Discussions}

In summary, we have presented and benchmarked the method of fTEDOPA and its generalizations to the finite-temperature case.
Our numerical examples illustrate the versatility of fTEDOPA in the study of fermionic open quantum systems and show the possible performance improvement by working in the Heisenberg picture.
Promising applications of fTEDOPA range across a variety of fields. 
Starting from quantum information processing like qubit implementations using quantum dots over to quantum thermodynamical processes with working systems coupled to fermionic reservoirs~\cite{Latta2011,Brantut2012}.

In particular, fTEDOPA could prove exceptionally useful in investigating phenomena in non-equilibrium quantum transport, as exemplified by the processes of energy and charge transfer through molecular junctions~\cite{schinabeck2016, Wang2011}.
A molecular junction is composed of at least one molecule coupled to two fermionic leads acting as source and drain.
Additionally, the molecular degrees of freedom couple to a single vibrational mode or a whole phononic bath.
One is then generally interested in the steady-state current through the junction as a function of the voltage applied to the leads.
Current approaches include quantum master equations have successfully simulated situations with weak molecule-lead coupling via a pertubative treatment~\cite{sowa2017,sowa2018}.
Another promising approach is that of hierarchical equations of motion, which works well for high temperatures and unstructured spectral densities~\cite{Suess2015, schinabeck2016}.
In contrast, fTEDOPA is expected to work well for arbitrary temperatures and makes no assumption on the system-bath interaction strength.
Furthermore, fTEDOPA allows for a controllable error in simulations.

One could speculate with the use of fTEDOPA to analyze exact models for charge transfer in molecular configurations that mimic reaction centers in actual photosynthetic systems~\cite{Grondelle1994} to provide insights on the combined effect of strong electron-phonon coupling and significant electronic delocalization.
Moreover, fermionic and bosonic TEDOPA could be combined to deal with situations where the system is subject to multiple, mixed environments such as an impurity in a Bose-Fermi mixture~\cite{Truscott2001,Schreck2001,Hadzibabic2002,Hadzibabic2003,Modugno2002}.

Although we have shown that working in the Heisenberg rather than in the Schr\"odinger picture is beneficial in some cases, a thorough analysis of the range of physical models where this holds is subject to future work.
Furthermore, comparing the performance of TEDOPA and thus also fTEDOPA to other DMRG-like methods is an important open problem.
Related methods differ in two aspects, the first being how the continuous bath is discretized, for instance using logarithmic or linear discretization. 
The effect of different spectral densities, including those that are highly structured, on the accuracy of different discretization warrants further study.

The second point of difference versus related methods is whether they rely on simulations performed in the discrete chain representation, analogous to TEDOPA, or in the discrete star representations, which are studied in Refs.~\cite{Wolf2014,Vega2015}.
Simulations in the chain picture are expected to benefit from the nearest-neighbor couplings, which is expected to keep correlations low for short simulation times.
On the other hand, the cost of initial state preparation could be lower for the star picture but interaction terms acting between distant sites could lead to large correlations in some situations that might make matrix product simulations costly.
Thus, differences in the overall performance of these two pictures for different spectral densities, interaction terms, temperatures and initial states is an open question.
Finally, an important open challenge is to devise methods for thermalized (f)TEDOPA for a system-environment interaction Hamiltonian that has no tensor-product form but instead comprises sums of more than one interaction term with a tensor product form.

\acknowledgements{
We thank Felipe Caycedo-Soler, Andisheh Khedri, Mark Mitchison, Andrea Smirne and Dario Tamascelli for helpful discussions.
MPS simulations are performed using the \texttt{mpnum} package~\cite{Suess2017a} by Daniel Suess and Milan Holzaepfel.
This work was supported by the ERC Synergy grant BioQ, the EU projects AsteriQs and Hyperdiamond, the EU QuantERA Project NanoSpin, Alexander von Humboldt Foundation via Humboldt Research Fellowship for Postdoctoral Researchers, University of Ulm via Forschungsbonus, DiaPol funded by BMBF (Federal Ministry of Education and Research), the state of Baden-W\"{u}rttemberg through bwHPC and the Deutsche Forschungsgemeinschaft (DFG) through grant no INST 40/467-1 FUGG (JUSTUS cluster).}

\bibliography{ftedopa}

\end{document}